\documentstyle[prl,epsf,psfrag,aps]{revtex}

\begin{document}

\title{Broadband teleportation}
\author{P.\ van Loock and Samuel L.\ Braunstein}
\address{Quantum Optics and Information Group,\\
School of Informatics, University of Wales, Bangor LL57 1UT, United Kingdom}
\author{H.\ J.\ Kimble}
\address{Norman Bridge Laboratory of Physics 12-33,\\ 
California Institute of Technology, Pasadena, California 91125}
\maketitle

\begin{abstract}
Quantum teleportation of an unknown broadband electromagnetic 
field is investigated. The continuous-variable teleportation
protocol by Braunstein and Kimble [Phys.\ Rev.\ Lett.\ 
{\bf 80}, 869 (1998)] for teleporting 
the quantum state of a single mode of the electromagnetic field is 
generalized for the case of a multimode field with finite bandwith.
We discuss criteria for continuous-variable teleportation with
various sets of input states and apply them to the teleportation
of broadband fields. We first consider as a set of input fields
(from which an independent state preparer draws the inputs to be
teleported) arbitrary pure Gaussian states with unknown coherent
amplitude (squeezed or coherent states).
This set of input states, further restricted to an alphabet of
coherent states, was used in the experiment by Furusawa {\it et al.}
[Science {\bf 282}, 706 (1998)]. It requires unit-gain teleportation
for optimizing the teleportation fidelity.
In our broadband scheme, the excess noise added through unit-gain 
teleportation due to the finite degree of the squeezed-state entanglement
is just twice the (entanglement) source's squeezing spectrum for its 
``quiet quadrature.'' The teleportation of one half of an
entangled state (two-mode squeezed vacuum state), i.e., 
``entanglement swapping,'' and its verification are optimized under
a certain nonunit gain condition. We will also give a broadband
description of this continuous-variable
entanglement swapping based on the single-mode scheme by
van Loock and Braunstein [Phys.\ Rev. A {\bf 61}, 10302 (2000)].
\end{abstract}

\section{Introduction}

Teleportation of an unknown quantum state is its disembodied transport 
through a classical channel, followed by its reconstitution, using the
quantum resource of entanglement. Quantum information cannot
be transmitted reliably via a classical channel alone, as this would 
allow us to replicate the classical signal and so produce copies of the
initial state, thus violating the no-cloning theorem \cite{Wootters}. 
More intuitively, any attempted measurement of the initial state only 
obtains partial information due to the Heisenberg uncertainty principle 
and the subsequently collapsed wave packet forbides information gain 
about the original state from further inspection. Attempts to circumvent 
this disability with more generalized measurements also fail \cite{Kraus}. 
 
Quantum teleportation was first proposed to transport an unknown state 
of any discrete quantum system, e.g., a spin-$\frac{1}{2}$ particle 
\cite{Benn}. In order to accomplish the teleportation, classical and 
quantum methods must go hand in hand. A part of the information encoded 
in the unknown input state is transmitted via the
quantum correlations between two separated subsystems in an entangled 
state shared by the sender and the receiver. In addition, classical 
information must be sent via a conventional channel. For the teleportation 
of a spin-$\frac{1}{2}$-particle state, the entangled state required 
is a pair of spins in a Bell state \cite{Sam2}. The classical information 
that has to be transmitted contains two bits in this case. 

Important steps toward the experimental implementation of quantum 
teleportation of single-photon polarization states have already 
been accomplished \cite{Bou,Mart}. However, a complete realization of
the original teleportation proposal \cite{Benn} has not been achieved
in these experiments, as either the state to be teleported is not 
independently coming from the outside \cite{Mart} or destructive detection
of the photons in the teleported state is employed as part of the protocol 
\cite{Bou}.
In the latter case, a teleported state did not emerge for subsequent
examination or exploitation. This situation has been termed ``a posteriori" 
teleportation, being accomplished via post selection of photoelectric
counting events \cite{Sam3}. Without post selection, the fidelity would not
have exceeded the value $\case{2}{3}$ required.

The teleportation of 
continuous quantum variables such as position and momentum of a particle 
\cite{Vaid} relies on the entanglement of the states in the original 
Einstein, Podolsky, and Rosen (EPR) paradox \cite{Einst}. 
In quantum optical terms, the observables analogous to the two conjugate
variables position and momentum of a particle are the quadrature 
amplitudes of a single mode of the electromagnetic field \cite{Walls}.
By considering the finite (nonsingular) degree of correlation between
these quadratures in a two-mode squeezed state \cite{Walls},
a realistic implementation for the teleportation of continuous 
quantum variables was proposed \cite{Sam}.
Based on this proposal, in fact, quantum teleportation of arbitrary
coherent states has been achieved with a fidelity $F=0.58\pm 0.02$
\cite{Furu}.
Without using entanglement, by purely classical communication, 
an average fidelity of 0.5 is the best that can be achieved
if the set of input states contains all coherent states \cite{Fuchs}.
The scheme with continuous quadrature amplitudes of a single 
mode enables an {\it ``a priori"} (or ``unconditional") teleportation with 
high efficiency \cite{Sam}, as reported in Refs.~\onlinecite{Sam4,Furu}.  
In this experiment, three criteria necessary for quantum teleportation were
achieved:

1. An unknown quantum state enters the sending station for teleportation.

2. A teleported state emerges from the receiving station for subsequent
evaluation or exploitation.

3. The degree of overlap between the input and the teleported states is
higher than that which could be achieved if the sending and the receiving
stations were linked only by a classical channel.

In continuous-variable teleportation, the teleportation process acts on an 
infinite-dimensional Hilbert space instead of the two-dimensional Hilbert 
space for the discrete spin variables. However, an arbitrary electromagnetic 
field has an infinite number of modes, or in other words, a finite bandwidth 
containing a continuum of modes. Thus, the teleportation of the quantum 
state of a broadband electromagnetic field requires the teleportation of a 
quantum state which is defined in the tensor product space of an infinite 
number of infinite-dimensional Hilbert spaces.
The aim of this paper is to extend the treatment of Ref.~\onlinecite{Sam}
to the case of a broadband field, and thereby to provide the theoretical 
foundation for laboratory investigations as in Refs.~\onlinecite{Sam4,Furu}.
In particular, we demonstrate that the two-mode squeezed state output of 
a nondegenerate optical parametric amplifier (NOPA) \cite{Ou} is a suitable 
EPR ingredient for the efficient teleportation of a broadband 
electromagnetic field. 

In the three above mentioned teleportation experiments, 
in Innsbruck \cite{Bou}, in Rome \cite{Mart}, and in Pasadena \cite{Furu}, 
the nonorthogonal input states to be teleported were 
single-photon polarization states (``qubits'') \cite{Bou,Mart} and 
coherent states \cite{Furu}.
From a true quantum teleportation device, however, we would also require
the capability of teleporting the entanglement source itself.
This teleportation of one half of an entangled state (``entanglement
swapping'' \cite{Zuk}) means to entangle two quantum systems
that have never directly interacted with each other. 
For discrete variables, a demonstration of entanglement swapping
with single photons has been reported by Pan {\it et al.} \cite{Pan}.
For continuous variables, experimental entanglement swapping has not yet
been realized in the laboratory, but there have been several theoretical
proposals of such an experiment.
Polkinghorne and Ralph \cite{Polk} suggested teleporting 
polarization-entangled states of single photons using squeezed-state 
entanglement where the output correlations are verified via Bell 
inequalities. Tan \cite{Tan} and van Loock and Braunstein \cite{PvL}
considered the unconditional teleportation (without post selection 
of ``successful'' events by photon detections) of one half of
a two-mode squeezed state using different protocols and verification.
Based on the single-mode scheme of Ref.~\onlinecite{PvL},
we will also present a broadband description of continuous-variable
entanglement swapping.    

\section{Teleportation of a single mode}

In the teleportation scheme of a single mode of the electromagnetic field
(for example, representing a single pulse or wave packet), 
the shared entanglement is a two-mode squeezed vacuum state \cite{Sam}. 
For infinite squeezing, this state contains exactly analogous 
quantum correlations as does the state described in the original EPR 
paradox, where the quadrature amplitudes of the two modes play the roles 
of position and momentum \cite{Sam}. The entangled state is sent in two
halves: one to ``Alice'' (the teleporter or sender) and the 
other one to ``Bob'' (the receiver), as illustrated in Fig.~1. 
In order to perform the teleportation, Alice has to couple the input 
mode she wants to teleport with her ``EPR mode'' at a beam splitter. 
The ``Bell detection'' of the $x$ quadrature at one beam splitter 
output, and of the $p$ quadrature at the other output, yields the 
classical results to be sent to Bob via a classical communication channel. 
In the limit of an infinitely squeezed EPR source, these classical results 
contain no information about the mode to be teleported. This is analogous 
to the Bell-state measurement of the spin-$\frac{1}{2}$-particle pair by 
Alice for the teleportation of a spin-$\frac{1}{2}$-particle state.  
The measured Bell state of the spin-$\frac{1}{2}$-particle pair determines 
whether the particles have equal or different spin projections. 
The spin projection of the individual particles, i.e., Alice's EPR particle
and her unknown input particle, remains completely unknown \cite{Benn}. 
According to this analogy, we call Alice's quadrature measurements for the 
teleportation of the state of a single mode (and of a multimode field in 
the following sections) ``Bell detection.''
Due to this Bell detection, the entanglement between Alice's ``EPR mode'' 
and Bob's ``EPR mode'' means that suitable phase-space displacements of 
Bob's mode convert it into a replica of Alice's unknown input mode 
(a perfect replica for infinite squeezing).
In order to perform these displacements, Bob needs the classical 
results of Alice's Bell measurement.
  
The previous protocol for the quantum teleportation of 
continuous variables used the Wigner distribution and its convolution 
formalism \cite{Sam}. The teleportation of a single mode 
of the electromagnetic field can also be recast in terms of Heisenberg 
equations for the quadrature amplitude operators, which is the formalism
that we employ in this paper. For that purpose, 
the Wigner function $W_{\rm EPR}$ describing the entangled state shared 
by Alice and Bob \cite{Sam} is replaced by equations for the 
quadrature amplitude operators of a two-mode squeezed vacuum state. Two 
independently squeezed vacuum modes can be described by \cite{Walls}
\begin{eqnarray}\label{1.1}
&&\hat{\bar{x}}_1=e^{r} \hat{\bar{x}}^{(0)}_1,\;\;\;\hat{\bar{p}}_1=
e^{-r} \hat{\bar{p}}^{(0)}_1,
\nonumber\\
&&\hat{\bar{x}}_2=e^{-r} \hat{\bar{x}}^{(0)}_2,\;\;\;\hat{\bar{p}}_2=
e^{r} \hat{\bar{p}}^{(0)}_2,
\end{eqnarray}
where a superscript `$(0)$' denotes initial vacuum modes and $r$ is the 
squeezing parameter.
Superimposing the two squeezed modes at a 50/50 beam splitter 
yields the two output modes
\begin{eqnarray}\label{1.2}
&&\hat{x}_1=\frac{1}{\sqrt{2}}e^{r} \hat{\bar{x}}^{(0)}_1
+\frac{1}{\sqrt{2}}e^{-r} \hat{\bar{x}}^{(0)}_2,\;\;\;
\hat{p}_1=\frac{1}{\sqrt{2}}e^{-r} \hat{\bar{p}}^{(0)}_1
+\frac{1}{\sqrt{2}}e^{r} \hat{\bar{p}}^{(0)}_2,\nonumber\\
&&\hat{x}_2=\frac{1}{\sqrt{2}}e^{r} \hat{\bar{x}}^{(0)}_1
-\frac{1}{\sqrt{2}}e^{-r} \hat{\bar{x}}^{(0)}_2,\;\;\;
\hat{p}_2=\frac{1}{\sqrt{2}}e^{-r} \hat{\bar{p}}^{(0)}_1
-\frac{1}{\sqrt{2}}e^{r} \hat{\bar{p}}^{(0)}_2.
\end{eqnarray}
The output modes 1 and 2 are now entangled to a finite degree in a 
two-mode squeezed vacuum state. In the limit of infinite squeezing, 
$r\to\infty$, both output modes become infinitely noisy, but also the 
EPR correlations between them become ideal:
$(\hat{x}_1-\hat{x}_2)\to 0$, $(\hat{p}_1+\hat{p}_2)\to 0$.
Now mode 1 is sent to Alice and mode 2 is sent to Bob. Alice's mode is 
then superimposed at a 50/50 beam splitter with the input mode 
``in'':
\begin{eqnarray}\label{1.3}
&&\hat{x}_{\rm u}=\frac{1}{\sqrt{2}}\hat{x}_{\rm in}-\frac{1}{\sqrt{2}}
\hat{x}_1,\;\;\;
\hat{p}_{\rm u}=\frac{1}{\sqrt{2}}\hat{p}_{\rm in}-\frac{1}{\sqrt{2}}
\hat{p}_1,\nonumber\\
&&\hat{x}_{\rm v}=\frac{1}{\sqrt{2}}\hat{x}_{\rm in}+\frac{1}{\sqrt{2}}
\hat{x}_1,\;\;\;
\hat{p}_{\rm v}=\frac{1}{\sqrt{2}}\hat{p}_{\rm in}+\frac{1}{\sqrt{2}}
\hat{p}_1.
\end{eqnarray}
Using Eqs.~(\ref{1.3}) we will find it useful to write Bob's mode 2 as
\begin{eqnarray}\label{mode2}
\hat{x}_2&=&\hat{x}_{\rm in}-(\hat{x}_1-\hat{x}_2)
-\sqrt{2}\hat{x}_{\rm u}\nonumber\\
&=&\hat{x}_{\rm in}-\sqrt{2}e^{-r} \hat{\bar{x}}^{(0)}_2
-\sqrt{2}\hat{x}_{\rm u},\nonumber\\
\hat{p}_2&=&\hat{p}_{\rm in}+(\hat{p}_1+\hat{p}_2)
-\sqrt{2}\hat{p}_{\rm v}\nonumber\\
&=&\hat{p}_{\rm in}+\sqrt{2}e^{-r} \hat{\bar{p}}^{(0)}_1
-\sqrt{2}\hat{p}_{\rm v}.
\end{eqnarray}
Alice's Bell detection yields certain classical values 
$x_{\rm u}$ and $p_{\rm v}$ for $\hat{x}_{\rm u}$ and $\hat{p}_{\rm v}$.
The quantum variables $\hat{x}_{\rm u}$ and $\hat{p}_{\rm v}$
become classically determined, random variables.
We indicate this by turning 
$\hat{x}_{\rm u}$ and $\hat{p}_{\rm v}$ into
$x_{\rm u}$ and $p_{\rm v}$.
The classical probability distribution of
$x_{\rm u}$ and $p_{\rm v}$ is associated with the quantum statistics
of the previous operators \cite{Sam}.
Now, due to the entanglement, Bob's mode 2 collapses into  
states that for $r\to\infty$ differ from Alice's input state 
only in (random) classical phase-space displacements. After receiving 
Alice's classical results $x_{\rm u}$ and $p_{\rm v}$, Bob displaces
his mode, 
\begin{eqnarray}\label{1.5}
\hat{x}_2\longrightarrow\hat{x}_{\rm tel}&=&\hat{x}_2+\Gamma\sqrt{2}
x_{\rm u},\nonumber\\
\hat{p}_2\longrightarrow\hat{p}_{\rm tel}&=&\hat{p}_2+\Gamma\sqrt{2}
p_{\rm v},
\end{eqnarray}
thus accomplishing the teleportation \cite{Sam}. The parameter $\Gamma$
describes a normalized gain for the transformation from classical
photocurrent to complex field amplitude. For $\Gamma=1$, 
Bob's displacement eliminates $x_{\rm u}$ and $p_{\rm v}$
appearing in Eqs.~(\ref{mode2}) after the collapse of $\hat{x}_{\rm u}$
and $\hat{p}_{\rm v}$ due to the Bell detection.
The teleported field then becomes
\begin{eqnarray}\label{1.6}
\hat{x}_{\rm tel}&=&\hat{x}_{\rm in}-\sqrt{2}e^{-r} 
\hat{\bar{x}}^{(0)}_2,
\nonumber\\
\hat{p}_{\rm tel}&=&\hat{p}_{\rm in}+\sqrt{2}e^{-r} 
\hat{\bar{p}}^{(0)}_1.
\end{eqnarray}
For an arbitrary gain $\Gamma$, we obtain
\begin{eqnarray}\label{gain}
\hat{x}_{\rm tel}&=&\Gamma\hat{x}_{\rm in}-\frac{\Gamma-1}{\sqrt{2}}
e^{r}\hat{\bar{x}}^{(0)}_1-\frac{\Gamma+1}{\sqrt{2}}e^{-r}
\hat{\bar{x}}^{(0)}_2,\nonumber\\
\hat{p}_{\rm tel}&=&\Gamma\hat{p}_{\rm in}+\frac{\Gamma-1}{\sqrt{2}}
e^{r}\hat{\bar{p}}^{(0)}_2+\frac{\Gamma+1}{\sqrt{2}}e^{-r}
\hat{\bar{p}}^{(0)}_1.
\end{eqnarray}
Note that these equations take no Bell detector inefficiencies into
account. 

Consider the case $\Gamma=1$. For infinite squeezing 
$r\to\infty$, Eqs.~(\ref{1.6}) describe perfect teleportation of the 
quantum state of the input mode. On the other hand, for the classical 
case of $r=0$, i.e., no squeezing and hence no entanglement, each of 
the teleported quadratures has {\it two} additional units of vacuum 
noise compared to the original input quadratures. These two units are 
so-called quantum duties or ``quduties" which have to be paid when 
crossing the border between quantum and classical domains \cite{Sam}.
The two quduties represent the minimal tariff for every 
``classical teleportation'' scheme \cite{Fuchs}.
One quduty, the unit of vacuum noise due to Alice's detection,
arises from her attempt to 
simultaneously measure the two conjugate variables $x_{\rm in}$ and 
$p_{\rm in}$ \cite{Arth}. This is the standard quantum limit for the 
detection of both quadratures \cite{Yama} when attempting to gain as 
much information as possible about the quantum state of a light field 
\cite{Leon}. 
The standard quantum limit yields a product of the measurement accuracies 
which is twice as large as the Heisenberg minimum uncertainty product. 
This product of the measurement accuracies contains the intrinsic quantum 
limit (Heisenberg uncertainty of the field to be detected) plus an additional 
unit of vacuum noise due to the detection \cite{Yama}. The second quduty 
arises when Bob uses the information of Alice's detection to generate the
state at amplitude $\sqrt{2}x_{\rm u}+i\sqrt{2}p_{\rm v}$ \cite{Sam}. It can 
be interpreted as the standard quantum limit imposed on state broadcasting.

\section{Teleportation criteria}

The teleportation scheme with Alice and Bob is complete 
without any further measurement. The quantum state teleported remains 
unknown to both Alice and Bob and need not be demolished in a detection 
by Bob as a final step.
However, maybe Alice and Bob are cheating. Instead of using an EPR 
channel, they try to get away without entanglement and use only a 
classical channel. In particular, for the realistic experimental
situation with finite squeezing and inefficient detectors where perfect
teleportation is unattainable, how may we verify that successful 
quantum teleportation has taken place?
To make this verification we shall introduce a third party, 
``Victor'' (the verifier), who is independent of Alice and Bob (Fig.~2). 
We assume that he prepares the initial input state
(drawn from a fixed set of states) and passes it on to 
Alice. After accomplishing the supposed teleportation, Bob sends the 
teleported state back to Victor. Victor's knowledge about the input state 
and detection of the teleported state enable Victor to verify
if quantum teleportation has really taken place.
For that purpose, however, Victor needs some measure that helps him to
assess when the similarity between the teleported state and the input state
exceeds a boundary that is only exceedable with entanglement.

\subsection{Teleporting Gaussian states with a coherent amplitude}

The single-mode teleportation scheme from Ref.~\onlinecite{Sam}
works for arbitrary input states, described by any Wigner function
$W_{\rm in}$. Teleporting states with a coherent amplitude 
as reliably as possible requires unit-gain teleportation (unit gain
in Bob's final displacement). Only in this case, the coherent 
amplitudes of the teleported mode always match those of the input mode
when Victor draws states with different amplitudes from the set
of input states in a sequence of trials. 
For this unit-gain teleportation, the teleported state $W_{\rm tel}$ 
is a convolution of the input $W_{\rm in}$ with a complex Gaussian of
variance $e^{-2r}$. Classical teleportation with $r=0$ then means
the teleported mode has an excess noise of two units of vacuum 
$\case{1}{2}+\case{1}{2}$ compared to the input, as also discussed in
the previous section. Any $r>0$ beats this classical scheme, i.e.,
if the input state is always recreated with the right amplitude and
less than two units of vacuum excess noise, we may call this already
quantum teleportation. Let us derive this result using the
least noisy model for classical communication.
For the input quadratures of Alice's sending station and the output
quadratures at Bob's receiving station, the least noisy (linear)
model if Alice and Bob are only classically communicating can be 
written as
\begin{eqnarray}\label{model}
\hat{x}_{\rm out,j}&=&\Gamma_x\,\hat{x}_{\rm in}+\Gamma_x\,
s_a^{-1}\hat{x}_a^{(0)}+s_{b,j}^{-1}\hat{x}^{(0)}_{b,j},\nonumber\\
\hat{p}_{\rm out,j}&=&\Gamma_p\,\hat{p}_{\rm in}-\Gamma_p\,
s_a\hat{p}_a^{(0)}+s_{b,j}\hat{p}^{(0)}_{b,j}.
\end{eqnarray}
This model takes into account that Alice and Bob can only 
communicate via classical signals, since arbitrarily many copies
of the output mode can be made by Bob where the subscript $j$
labels the $j$th copy. In addition, it ensures that the output
quadratures satisfy the commutation relations
\begin{eqnarray}\label{commut}
[\hat{x}_{\rm out,j},\hat{p}_{\rm out,k}]&=&(i/2)\;\delta_{jk},\nonumber\\
\left[\hat{x}_{\rm out,j},\hat{x}_{\rm out,k}\right]&=&[\hat{p}_{\rm out,j},
\hat{p}_{\rm out,k}]=0\,\,\,.
\end{eqnarray}
Since we are only interested in one single copy of the output we drop
the label $j$.
The parameter $s_a$ is given by Alice's measurement strategy
and determines the noise penalty due to her homodyne detections. 
The gains $\Gamma_x$ and $\Gamma_p$ can be manipulated by Bob as well
as the parameter $s_b$ determining the noise distribution of
Bob's original mode. The set of input states may contain pure Gaussian
states with a coherent amplitude, described by
$\hat{x}_{\rm in}=\langle\hat{x}_{\rm in}\rangle+s_v^{-1}\hat{x}^{(0)}$
and $\hat{p}_{\rm in}=
\langle\hat{p}_{\rm in}\rangle+s_v\hat{p}^{(0)}$,
where Victor can choose in each trial the coherent amplitude
and if and to what extent the input is squeezed (parameter $s_v$). 
Since Bob always wants to reproduce the input amplitude,
he is restricted to unit gain, symmetric in both quadratures,
$\Gamma_x=\Gamma_p=1$. First, after obtaining the output states from Bob,
Victor verifies if their amplitudes match the corresponding input
amplitudes. If not, all the following considerations concerning the
excess noise are redundant, because Alice and Bob can always
manipulate this noise by fiddling the gain (less than unit gain reduces
the excess noise). If Victor finds overlapping amplitudes in all trials
(at least within some error range),
he looks at the excess noise in each trial.
For that purpose, let us define the normalized variance
\begin{eqnarray}\label{telin}
V_{\rm out,in}^{\hat{x}}&\equiv&\frac{\langle\Delta(\hat{x}_{\rm out}-
\hat{x}_{\rm in})^2\rangle}{\langle\Delta\hat{x}^2
\rangle_{\rm vacuum}}\,\,\,,
\end{eqnarray}
and analogously $V_{\rm out,in}^{\hat{p}}$ with 
$\hat{x}\rightarrow\hat{p}$ throughout
$[\langle\Delta\hat{o}^2\rangle\equiv{\rm var}(\hat{o})]$. 
Using Eqs.~(\ref{model}) with unit gain, we obtain the product
\begin{eqnarray}\label{product}
V_{\rm out,in}^{\hat{x}}V_{\rm out,in}^{\hat{p}}&=&
(s_a^{-2}+s_b^{-2})(s_a^2+s_b^2).
\end{eqnarray}
It is minimized for $s_a=s_b$, yielding
$V_{\rm out,in}^{\hat{x}}V_{\rm out,in}^{\hat{p}}=4$.
The optimum value of 4 is exactly the result we obtain for what we may 
call classical teleportation,
$V_{\rm tel,in}^{\hat{x}}(r=0)V_{\rm tel,in}^{\hat{p}}(r=0)=4$,
using Eqs.~(\ref{1.6}) with subscript `out' $\rightarrow$ `tel'
in Eq.~(\ref{telin}). Thus, we can write our first ``fundamental''
limit for teleporting states with a coherent amplitude as 
\begin{eqnarray}\label{limit}
V_{\rm out,in}^{\hat{x}}V_{\rm out,in}^{\hat{p}}&\geq&
V_{\rm tel,in}^{\hat{x}}(r=0)V_{\rm tel,in}^{\hat{p}}(r=0)=4\,\,\,.
\end{eqnarray}
If Victor, comparing the output states with the input states, always 
finds violations of this inequality, he may already have big
confidence in Alice's and Bob's honesty (i.e., that they indeed have
used entanglement). Equation~(\ref{limit}) may also enable us already
to assess if a scheme or protocol is capable of quantum teleportation.
Alternatively, instead of looking at the products
$V_{\rm out,in}^{\hat{x}}V_{\rm out,in}^{\hat{p}}$, we could also 
use the sums $V_{\rm out,in}^{\hat{x}}+V_{\rm out,in}^{\hat{p}}=
s_a^{-2}+s_b^{-2}+s_a^2+s_b^2$ that are minimized for $s_a=s_b=1$.
Then we find the classical boundary
$V_{\rm out,in}^{\hat{x}}+V_{\rm out,in}^{\hat{p}}\geq 4$.   

However, taking into account all the assumptions made for the
derivation of Eq.~(\ref{limit}), this boundary appears to be less 
fundamental. First, we have only assumed a linear model.
Secondly, we have only considered the variances of two conjugate 
observables and a certain kind of measurement of these.
An entirely rigorous criterion for quantum teleportation
should take into account all possible variables, 
measurements and strategies that can be used by Alice and Bob.
Another ``problem'' of our boundary Eq.~(\ref{limit}) is
that the variances $V_{\rm out,in}$ are not directly measurable,
because the input state is destroyed by the teleportation process.
However, for Gaussian input states, Victor can combine his
knowledge of the input variances $V_{\rm in}$ with the detected 
variances $V_{\rm out}$ in order to infer $V_{\rm out,in}$.
With a more specific set of Gaussian input states, 
namely coherent states, the least noisy model for classical 
communication allows us to determine the directly measurable
``fundamental'' limit for the normalized variances of the 
output states
\begin{eqnarray}\label{limit2}
V_{\rm out}^{\hat{x}}V_{\rm out}^{\hat{p}}\geq 9\,\,\,.
\end{eqnarray} 
But still we need to bear in mind that we did not consider 
all possible strategies of Alice and Bob.
Also for arbitrary $s_v$ (set of input states contains all
coherent and squeezed states), Eq.~(\ref{limit2}) represents a classical
boundary, as
\begin{eqnarray}\label{product2}
V_{\rm out}^{\hat{x}}V_{\rm out}^{\hat{p}}&=&
(s_v^{-2}+s_a^{-2}+s_b^{-2})(s_v^2+s_a^2+s_b^2)
\end{eqnarray}
is minimized for $s_v=s_a=s_b$, yielding 
$V_{\rm out}^{\hat{x}}V_{\rm out}^{\hat{p}}=9$.
However, since $s_v$ is unknown to Alice and Bob in every trial,
they can attain this classical minimum only by accident.
For $s_v$ fixed, e.g., $s_v=1$ (set of input states contains ``only''
coherent states), Alice and Bob knowing this $s_v$ can always
satisfy $V_{\rm out}^{\hat{x}}V_{\rm out}^{\hat{p}}=9$ in the classical
model. Alternatively, the sums $V_{\rm out}^{\hat{x}}+
V_{\rm out}^{\hat{p}}=s_v^{-2}+s_a^{-2}+s_b^{-2}+s_v^2+s_a^2+s_b^2$
are minimized with $s_a=s_b=1$. In this case, we obtain the 
$s_v$-dependent boundary $V_{\rm out}^{\hat{x}}+
V_{\rm out}^{\hat{p}}\geq s_v^{-2}+s_v^2+4$. Without knowing
$s_v$, Alice and Bob can always attain this minimum in the classical
model. In every trial, Victor must combine his knowledge of $s_v$
with the detected output variances in order to find violations
of this sum inequality.

Ralph and Lam \cite{Ralph} define the classical boundaries
\begin{eqnarray}\label{ralph1}
V_{\rm c}^{\hat{x}}+V_{\rm c}^{\hat{p}}\geq 2
\end{eqnarray}
and
\begin{eqnarray}\label{ralph2}
T_{\rm out}^{\hat{x}}+T_{\rm out}^{\hat{p}}\leq 1\,\,\,,
\end{eqnarray}
using the conditional variance
\begin{eqnarray}\label{cond}
V_{\rm c}^{\hat{x}}&\equiv&\frac{\langle\Delta\hat{x}^2_{\rm out}
\rangle}{\langle\Delta\hat{x}^2\rangle_{\rm vacuum}}
\left(1-\frac{|\langle\Delta\hat{x}_{\rm out}\Delta\hat{x}_{\rm in}
\rangle|^2}{\langle\Delta\hat{x}^2_{\rm out}\rangle\langle
\Delta\hat{x}^2_{\rm in}\rangle}\right),
\end{eqnarray}
and analogously for $V_{\rm c}^{\hat{p}}$ with 
$\hat{x}\rightarrow\hat{p}$ throughout,
and the transfer coefficient
\begin{eqnarray}\label{trans1}
T_{\rm out}^{\hat{x}}&\equiv&\frac{{\rm SNR}_{\rm out}^{\hat{x}}}{{\rm SNR}
_{\rm in}^{\hat{x}}}\,\,\,,
\end{eqnarray}
and analogously $T_{\rm out}^{\hat{p}}$ with $\hat{x}\rightarrow\hat{p}$ 
throughout. Here, SNR denotes the signal to noise ratio for the square of 
the mean amplitudes, namely ${\rm SNR}_{\rm out}^{\hat{x}}=\langle
\hat{x}_{\rm out}\rangle^2/\langle\Delta\hat{x}_{\rm out}^2\rangle$. 

Alice and Bob using only classical communication are not able to 
violate {\it either} of the two inequalities 
Eq.~(\ref{ralph1}) and Eq.~(\ref{ralph2}).
In fact, these boundaries are two independent limits, 
each of them unexceedable in a classical scheme.
However, Alice and Bob can simultaneously approach 
$V_{\rm c}^{\hat{x}}+V_{\rm c}^{\hat{p}}=2$ and 
$T_{\rm out}^{\hat{x}}+T_{\rm out}^{\hat{p}}=1$ using either
an asymmetric classical detection and transmission scheme
with coherent-state inputs or a symmetric classical scheme with 
squeezed-state inputs \cite{Ralph}.
For quantum teleportation, Ralph and Lam \cite{Ralph} require their 
classical limits be simultaneously exceeded, $V_{\rm c}^{\hat{x}}+
V_{\rm c}^{\hat{p}}<2$ and $T_{\rm out}^{\hat{x}}+T_{\rm out}^{\hat{p}}>1$.
This is only possible using more than 3 dB squeezing in the entanglement 
source \cite{Ralph}.  
Apparently, these criteria determine a classical boundary different from
ours in Eq.~(\ref{limit}). For example, in unit-gain teleportation,
our inequality Eq.~(\ref{limit}) is violated for any nonzero squeezing 
$r>0$. Let us briefly explain why we encounter this discrepancy.
We have a priori assumed unit gain in our scheme to achieve outputs
and inputs overlapping in their mean values. This assumption is, of course,
motivated by the assessment that good teleportation means good similarity
between input and output {\it states} (here, to be honest, we already have
something in mind similar to the fidelity, introduced in the next section).
First, Victor has to check the match of the amplitudes before looking
at the variances. Ralph and Lam permit arbitrary gain,
because they are not interested in the similarity of input and output
{\it states}, but in certain correlations that manifest
separately in the individual quadratures \cite{Ralph2}.
This point of view originates from the context of quantum nondemolition
(QND) measurements \cite{Brag}, which are focused on a single QND variable 
while the conjugate variable is not of interest.
For arbitrary gain, an inequality as in Eq.~(\ref{ralph2}),
containing the input and output mean values, has to be added to an 
inequality only for variances as in Eq.~(\ref{ralph1}).
Ralph and Lam's {\it best} classical protocol permits output states 
completely different from the input states, e.g., via asymmetric detection
where the lack of information in one quadrature leads on average 
to output states with amplitudes completely different from the input
states. The asymmetric scheme means that Alice is {\it not} 
attempting to gain as much information about the {\it quantum state}
as possible, as in an Arthurs-Kelly measurement \cite{Arth}. 
The Arthurs-Kelly measurement, 
however, is exactly what Alice should do in our {\it best} classical
protocol, i.e., classical teleportation.  
Therefore, our best classical protocol always achieves output states
already pretty similar to the input states.
Apparently, ``the best'' that can be classically achieved 
has a different meaning from Ralph and Lam's point of view and
from ours. Then it is no surprise that the classical boundaries 
differ as well. 
Apart from these differences, however, Ralph and Lam's criteria do
have something in common with our criterion given by Eq.~(\ref{limit}):
they also do not satisfy the rigor
we require from criteria for quantum teleportation taking into account
everything Alice and Bob can do. By limiting the set of input states
to coherent states, we are able to present such a rigorous criterion
in the next section. 

\subsection{The fidelity criterion for coherent-state teleportation}

The rigorous criterion we are looking for to determine the best classical
teleportation and to quantify the distinction between classical and 
quantum teleportation relies on the fidelity $F$, for an arbitrary input 
state $|\psi_{\rm in}\rangle$ defined by \cite{Fuchs}
\begin{eqnarray}\label{fid1}
F\equiv\langle\psi_{\rm in}|\hat{\rho}_{\rm out}|
\psi_{\rm in}\rangle.
\end{eqnarray}
It is an excellent measure for the similarity between the input 
and the output state and equals one only if 
$\hat{\rho}_{\rm out}=|\psi_{\rm in}\rangle\langle\psi_{\rm in}|$.
Now Alice and Bob know that Victor draws his states
$|\psi_{\rm in}\rangle$ from a fixed set, but they do not know 
which particular state is drawn in a single trial.
Therefore, an average fidelity should be considered \cite{Fuchs},
\begin{eqnarray}\label{fid2}
F_{\rm av}=\int P(|\psi_{\rm in}\rangle)
\langle\psi_{\rm in}|\hat{\rho}_{\rm out}|
\psi_{\rm in}\rangle d|\psi_{\rm in}\rangle,
\end{eqnarray}
where $P(|\psi_{\rm in}\rangle)$ is the probability of drawing
a particular state $|\psi_{\rm in}\rangle$, and the integral runs
over the entire set of input states.
If the set of input states contains simply all possible quantum states 
in an infinite-dimensional Hilbert space (i.e., the input state is
completely unknown apart from the Hilbert-space dimension), the 
best average fidelity achievable without entanglement is zero.
If the set of input states is restricted to coherent states of amplitude
$\alpha_{\rm in}=x_{\rm in}+ip_{\rm in}$ and
$F=\langle\alpha_{\rm in}|\hat{\rho}_{\rm out}|\alpha_{\rm in}\rangle$,
on average, the fidelity achievable in a purely classical scheme 
(when averaged across the entire complex plane) is bounded by 
\cite{Fuchs}
\begin{eqnarray}\label{fid3}
F_{\rm av}\leq\frac{1}{2}\,\,\,.
\end{eqnarray}
Let us illustrate these nontrivial results with our single-mode
teleportation equations.
Up to a factor $\pi$, the fidelity 
$F=\langle\alpha_{\rm in}|\hat{\rho}_{\rm tel}|\alpha_{\rm in}\rangle$
is the $Q$ function of the teleported mode evaluated for 
$\alpha_{\rm in}$:
\begin{eqnarray}\label{fid4}
F=\pi Q_{\rm tel}(\alpha_{\rm in})=\frac{1}{2\sqrt{\sigma_x\sigma_p}}
\exp\left[-(1-\Gamma)^2\left(\frac{x_{\rm in}^2}{2\sigma_x}
+\frac{p_{\rm in}^2}{2\sigma_p}\right)\right], 
\end{eqnarray}
where $\Gamma$ is the gain from the previous sections and $\sigma_x$ 
and $\sigma_p$ are the variances of the
Q function of the teleported mode for the corresponding quadratures. 
These variances are according to Eqs.~(\ref{gain}) for a coherent-state
input and $\langle\Delta\hat{x}^2\rangle_{\rm vacuum}
=\langle\Delta\hat{p}^2\rangle_{\rm vacuum}=\case{1}{4}$ given by 
\begin{eqnarray}\label{fid5}
\sigma_x=\sigma_p=
\frac{1}{4}(1+\Gamma^2)+\frac{e^{2r}}{8}(\Gamma-1)^2+
\frac{e^{-2r}}{8}(\Gamma+1)^2.
\end{eqnarray}
For classical teleportation ($r=0$) and $\Gamma=1$, we obtain 
$\sigma_x=\sigma_p=\case{1}{2}+\case{1}{4}V_{\rm tel,in}^{\hat{x}}(r=0)=
\case{1}{2}+\case{1}{4}V_{\rm tel,in}^{\hat{p}}(r=0)=\case{1}{2}+
\case{1}{2}=1$ and indeed $F=F_{\rm av}=\case{1}{2}$.
In order to obtain a better fidelity, entanglement is necessary. 
Then, if $\Gamma=1$, we obtain $F=F_{\rm av}>\case{1}{2}$ for any $r>0$.
For $r=0$, the fidelity drops to zero as 
$\Gamma\to\infty$ since the mean amplitude of the teleported state does not 
match that of the input state and the excess noise increases. 
For $r=0$ and $\Gamma=0$, the fidelity becomes 
$F=\exp(-|\alpha_{\rm in}|^2)$. Upon averaging over all possible 
coherent-state inputs, this fidelity also vanishes. 
Assuming nonunit gain, it is crucial to consider the average
fidelity $F_{\rm av}\neq F$. When averaging across the entire complex 
plane, any nonunit gain yields $F_{\rm av}=0$. 
This is exactly why Victor should first check the match of the 
amplitudes for different input states. If Alice and Bob are cheating
and fiddle the gain in a classical scheme, a sufficiently large input
amplitude reveals the truth.
These considerations also apply to the asymmetric classical detection
and transmission scheme with a coherent-state input 
\cite{Ralph} discussed in the previous section.
Of course, the asymmetric scheme does not provide an improvement in the 
fidelity. In fact, the average fidelity drops to zero, if Alice
detects only one quadrature (and gains complete information about this 
quadrature) and Bob obtains the full information about the measured 
quadrature, but no information about the second quadrature.
In an asymmetric classical scheme, Alice and Bob stay far within 
the classical domain $F_{\rm av}<\case{1}{2}$. The best classical scheme 
with respect to the fidelity is the symmetric one (``classical
teleportation'') with $F_{\rm av}=\case{1}{2}$.

The supposed limitation of the fidelity
criterion that the set of input states contains ``only'' coherent
states is compensated by having an entirely rigorous criterion.
Of course, the fidelity criterion does not limit the possible
input states for which the presented protocol works.
It does not mean we can only teleport coherent states (as we will clearly
see in the next section). However, so far, it is the only criterion
that enables the experimentalist to rigorously verify quantum 
teleportation. That is why Furusawa et al. \cite{Furu} were happy
to have used coherent-state inputs, because they could rely on 
a strict and rigorous criterion (and not only because coherent states
are the most readily available source for the state preparer Victor).

\subsection{Teleporting entangled states: entanglement swapping}

From a true quantum teleportation device, 
we require that it can not only teleport
nonorthogonal states very similar to classical states (such as coherent 
states), but also extremely nonclassical states such as entangled states.
When teleporting one half of an entangled state (``entanglement 
swapping''), we are certainly much more interested in the 
preservation of the inseparability than in the match of any input 
and output amplitudes. We can say that entanglement swapping is
successful, if the initially unentangled modes become entangled
via the teleportation process (even, if this is accompanied by a decrease
of the quality of the initial entanglement).
In Ref.~\onlinecite{PvL} has been shown, that the single-mode 
teleportation scheme enables entanglement swapping
for any nonzero squeezing ($r>0$) in the two initial entangled states
(of which one provides the teleporter's input and the other one the
EPR channel or vice versa). 

Let us introduce ``Claire'' who performs the Bell detection of modes 2 and
3 (Fig.~3). Before her measurement, mode 1 (Alice's mode) is entangled
with mode 2, and mode 3 is entangled with mode 4 (Bob's mode) \cite{PvL}.
Due to Claire's detection, mode 1 and 4 are projected on entangled
states. Entanglement is teleported in every single projection 
(for every measured value of $x_{\rm u}$ and $p_{\rm v}$) without
any further local displacement \cite{PvL3}. 
How can we verify that entanglement
swapping was successful? Simply, by verifying that Alice and Bob,
who initially did not share any entanglement, are able to perform
quantum teleportation using mode 1 and 4 after entanglement swapping 
\cite{PvL}. But then we urgently need a rigorous criterion for
quantum teleportation that unambigously recognizes when Alice and
Bob have used entanglement and when they have not.
Now, again, we can rely on the fidelity criterion for coherent-state
teleportation. Alice and Bob again have to convince Victor that they
are using entanglement and are not cheating. Of course, this is only a
reliable verification scheme of entanglement swapping, if one can be 
sure that Alice and Bob did not share entanglement prior to entanglement
swapping and that Claire is not allowed to perform unit-gain
displacements (or that Claire is not allowed to receive any classical
information). Otherwise, Victor's coherent-state input could be
teleported step by step from Alice to Claire (with unit gain) and
from Claire to Bob (with unit gain). This protocol, however, requires
more than 3 dB squeezing in both entanglement sources (if equally 
squeezed) to ensure $F_{\rm av}>\case{1}{2}$ \cite{PvL}. 
Using entanglement swapping, Alice and Bob can achieve 
$F_{\rm av}>\case{1}{2}$ for any squeezing, but one of them has to 
perform local displacements based on Claire's measurement results. 
Any gain is allowed in these displacements, since in entanglement
swapping, we are not interested
in the transfer of coherent amplitudes (and the two initial two-mode
squeezed states are vacuum states anyway).  
But only the optimum gain $\Gamma_{\rm swap}=\tanh 2r$ ensures 
$F_{\rm av}>\case{1}{2}$ for any squeezing and provides the optimum
fidelity \cite{PvL}.
Unit gain $\Gamma_{\rm swap}=1$ in entanglement swapping would
require more than 3 dB squeezing in both entanglement sources
(if equally squeezed) to achieve $F_{\rm av}>\case{1}{2}$ \cite{PvL},
or to confirm the teleportation of entanglement via detection of the
combined entangled modes \cite{Tan}.

We will also give a broadband protocol of entanglement swapping as a 
``nonunit-gain teleportation.'' The verification of entanglement
swapping via the fidelity criterion for coherent-state teleportation
demonstrates how useful this criterion is. Less rigorous criteria, 
as presented in Sec. III A, cannot reliably tell us if Alice
and Bob use entanglement emerging from entanglement swapping.
Furthermore, the entanglement swapping scheme demonstrates that
a two-mode squeezed state enables {\it true} quantum teleportation 
for any nonzero squeezing. Requiring more than 3 dB squeezing,
as it is necessary for quantum teleportation according to
Ralph and Lam \cite{Ralph}, is not necessary for the teleporation
of entanglement.

\section{Broadband entanglement}

In this section, we demonstrate that the EPR state required for
broadband teleportation can be generated either directly by
nondegenerate parametric down conversion or by combining
two independently squeezed fields produced via degenerate down
conversion or any other nonlinear interaction. 

First, we review the results of Ref.~\onlinecite{Ou} based on the 
input-output formalism of Collett and Gardiner \cite{Coll} where a 
nondegenerate optical parametric amplifier in a cavity (NOPA) is 
studied. We will see that the upper and lower sidebands of the NOPA 
output have correlations similar to those of the two-mode squeezed state in 
Eqs.~(\ref{1.2}). The optical parametric oscillator is considered 
polarization nondegenerate but frequency ``degenerate'' (equal 
center frequency for the orthogonally polarized output modes).
The interaction between the two modes is due to the nonlinear $\chi^{(2)}$ 
medium (in a cavity) and may be described by the interaction Hamiltonian
\begin{eqnarray}\label{1.8}
\hat{H}_{I}=i\hbar\kappa(\hat{a}^{\dagger}_1\hat{a}^{\dagger}_2 
e^{-2i\omega_0t}-\hat{a}_1\hat{a}_2 e^{2i\omega_0t}).
\end{eqnarray}
The undepleted pump field amplitude at frequency $2\omega_0$ is described 
as a {\it c} number and has been absorbed into the coupling $\kappa$ which 
also contains the $\chi^{(2)}$ susceptibility. Without loss of generality 
$\kappa$ can be taken to be real. The dynamics of the two cavity modes 
$\hat{a}_1$ and $\hat{a}_2$ are governed by the above interaction 
Hamiltonian, and input-output relations can be derived relating the 
cavity modes to the external vacuum input modes $\hat{b}^{(0)}_1$ 
and $\hat{b}^{(0)}_2$, the external output modes $\hat{b}_1$ and 
$\hat{b}_2$, and two unwanted vacuum modes $\hat{c}^{(0)}_1$ and 
$\hat{c}^{(0)}_2$ describing cavity losses (Fig.~4). Recall, the 
superscript `$(0)$' refers to vacuum modes. We define uppercase 
operators in the rotating frame about the center frequency $\omega_0$, 
\begin{eqnarray}\label{1.9}
\hat{O}(t)=\hat{o}(t)e^{i\omega_0t},
\end{eqnarray}
with $\hat{O}=[\hat{A}_{1,2};\hat{B}_{1,2};\hat{B}^{(0)}_{1,2};
\hat{C}^{(0)}_{1,2}]$ and the full Heisenberg operators $\hat{o}=
[\hat{a}_{1,2};\hat{b}_{1,2};\hat{b}^{(0)}_{1,2};\hat{c}^{(0)}_{1,2}]$. 
By the Fourier transformation
\begin{eqnarray}\label{1.10}
\hat{O}(\Omega)=\frac{1}{\sqrt{2\pi}}\int dt\,\hat{O}(t)e^{i\Omega t},
\end{eqnarray}
the fields are now described as functions of the modulation frequency 
$\Omega$ with commutation relation $[\hat{O}(\Omega),\hat{O}^{\dagger}
(\Omega')]=\delta(\Omega - \Omega')$ for $\hat{B}_{1,2}$, 
$\hat{B}^{(0)}_{1,2}$ and $\hat{C}^{(0)}_{1,2}$ since $[\hat{O}(t),
\hat{O}^{\dagger}(t')]=\delta(t - t')$. Expressing the outgoing modes 
in terms of the incoming vacuum modes, one obtains \cite{Ou}
\begin{eqnarray}\label{1.11}
\hat{B}_j(\Omega)=G(\Omega)\hat{B}^{(0)}_j(\Omega)+g(\Omega)
\hat{B}^{(0)\dagger}_k(-\Omega)
+\bar{G}(\Omega)\hat{C}^{(0)}_j(\Omega)+\bar{g}(\Omega)\hat{C}
^{(0)\dagger}_k(-\Omega),
\end{eqnarray}
where $k=3-j$, $j=1,2$ (so $k$ refers to the opposite mode to $j$), 
and with coefficients to be specified later.
The two cavity modes have been assumed to be both on resonance with 
half the pump frequency at $\omega_0$. 

Let us investigate the lossless case where the output fields become 
\begin{eqnarray}\label{1.12}
\hat{B}_j(\Omega)=G(\Omega)\hat{B}^{(0)}_j(\Omega)+g(\Omega)\hat{B}^{(0)
\dagger}_k(-\Omega),
\end{eqnarray}
with the functions $G(\Omega)$ and $g(\Omega)$ of Eq.~(\ref{1.11}) 
simplifying to 
\begin{eqnarray}\label{1.13}
G(\Omega)&=&\frac{\kappa^2+\gamma^2/4+\Omega^2}{(\gamma/2-i 
\Omega)^2-\kappa^2}\,\,\,,\nonumber\\
g(\Omega)&=&\frac{\kappa\gamma}{(\gamma/2-i \Omega)^2-\kappa^2}\,\,\,.
\end{eqnarray}
Here, the parameter $\gamma$ is a damping rate of the cavity 
(Fig.~4) and is assumed to be equal for both polarizations. 
Equation (\ref{1.12}) represents the input-output relations for a 
lossless NOPA.

Following Ref.~\onlinecite{Schum}, we introduce frequency resolved
quadrature amplitudes given by 
\begin{eqnarray}\label{1.14}
\hat{X}_j(\Omega)&=&\case{1}{2}[\hat{B}_j(\Omega)+\hat{B}^{\dagger}_j
(-\Omega)],\nonumber\\
\hat{P}_j(\Omega)&=&\case{1}{2i}[\hat{B}_j(\Omega)-\hat{B}^{\dagger}_j
(-\Omega)],\nonumber\\
\hat{X}^{(0)}_j(\Omega)&=&\case{1}{2}[\hat{B}^{(0)}_j(\Omega)+\hat{B}
^{(0)\dagger}_j(-\Omega)],\nonumber\\
\hat{P}^{(0)}_j(\Omega)&=&\case{1}{2i}[\hat{B}^{(0)}_j(\Omega)-\hat{B}
^{(0)\dagger}_j(-\Omega)],
\end{eqnarray}
provided $\Omega\ll \omega_0$. Using them Eq.~(\ref{1.12}) becomes
\begin{eqnarray}\label{1.15}
\hat{X}_j(\Omega)&=&G(\Omega)\hat{X}^{(0)}_j(\Omega)+g(\Omega)
\hat{X}^{(0)}_k(\Omega),\nonumber\\
\hat{P}_j(\Omega)&=&G(\Omega)\hat{P}^{(0)}_j(\Omega)-g(\Omega)
\hat{P}^{(0)}_k(\Omega).
\end{eqnarray}
Here, we have used $G(\Omega)=G^*(-\Omega)$ and $g(\Omega)=g^*(-\Omega)$.

At this juncture, we show that the output quadratures of a lossless NOPA 
in Eqs.~(\ref{1.15}) correspond to two independently squeezed modes 
coupled to a two-mode squeezed state at a beam splitter. 
The operational significance of this fact is that
the EPR state required for broadband teleportation can be created
either by nondegenerate parametric down conversion as described by the
interaction Hamiltonian in Eq.~(\ref{1.8}), or by combining at a 
beam splitter two independently squeezed fields generated via degenerate 
down conversion \cite{Kimb} (as done in the teleportation
experiment of Ref.~\onlinecite{Furu}). 

Let us thus define the superpositions of the two output modes
(barred quantities)  
\begin{eqnarray}\label{1.16}
\hat{\bar{B}}_1&\equiv&\case{1}{\sqrt{2}}(\hat{B}_1+\hat{B}_2),
\nonumber\\
\hat{\bar{B}}_2&\equiv&\case{1}{\sqrt{2}}(\hat{B}_1-\hat{B}_2),
\end{eqnarray}
and of the two vacuum input modes
\begin{eqnarray}\label{1.17}
\hat{\bar{B}}^{(0)}_1&\equiv&\case{1}{\sqrt{2}}(\hat{B}^{(0)}_1+
\hat{B}^{(0)}_2),\nonumber\\
\hat{\bar{B}}^{(0)}_2&\equiv&\case{1}{\sqrt{2}}(\hat{B}^{(0)}_1-
\hat{B}^{(0)}_2).
\end{eqnarray}
In terms of these superpositions, Eq.~(\ref{1.12}) becomes
\begin{eqnarray}\label{1.18}
\hat{\bar{B}}_1(\Omega)&=&G(\Omega)\hat{\bar{B}}^{(0)}_1(\Omega)+
g(\Omega)\hat{\bar{B}}^{(0)\dagger}_1
(-\Omega),\nonumber\\
\hat{\bar{B}}_2(\Omega)&=&G(\Omega)\hat{\bar{B}}^{(0)}_2(\Omega)-
g(\Omega)\hat{\bar{B}}^{(0)\dagger}_2
(-\Omega).
\end{eqnarray}
In Eqs.~(\ref{1.18}), the initially coupled modes of Eq.~(\ref{1.12}) 
are decoupled, corresponding to two independent degenerate parametric
amplifiers.

In the limit $\Omega\to 0$, the two modes of Eqs.~(\ref{1.18}) are 
each in the same single-mode squeezed state as the two modes in 
Eqs.~(\ref{1.1}). More explicitly, by setting $G(0)=\cosh r$ and 
$g(0)=\sinh r$, the annihilation operators
\begin{eqnarray}\label{1.19}
\hat{\bar{B}}_1&=&\cosh r\hat{\bar{B}}^{(0)}_1+\sinh r\hat{\bar{B}}^{(0)
\dagger}_1,\nonumber\\
\hat{\bar{B}}_2&=&\cosh r\hat{\bar{B}}^{(0)}_2-\sinh r\hat{\bar{B}}^{(0)
\dagger}_2,
\end{eqnarray}
have the quadrature operators
\begin{eqnarray}\label{1.20}
&&\hat{\bar{X}}_1=e^{r} \hat{\bar{X}}^{(0)}_1,\;\;\;\hat{\bar{P}}_1=
e^{-r} \hat{\bar{P}}^{(0)}_1,
\nonumber\\
&&\hat{\bar{X}}_2=e^{-r} \hat{\bar{X}}^{(0)}_2,\;\;\;\hat{\bar{P}}_2=
e^{r} \hat{\bar{P}}^{(0)}_2.
\end{eqnarray}
From the alternative perspective of superimposing two independently 
squeezed modes at a 50/50 beam splitter to obtain the EPR state, we must 
simply invert the transformation of Eqs.~(\ref{1.16}) and recouple the two 
modes:
\begin{eqnarray}\label{1.21}
\hat{B}_1&=&\case{1}{\sqrt{2}}(\hat{\bar{B}}_1+\hat{\bar{B}}_2)\nonumber\\
&=&\case{1}{\sqrt{2}}[\cosh r(\hat{\bar{B}}^{(0)}_1+\hat{\bar{B}}^{(0)}_2)
+\sinh r(\hat{\bar{B}}^{(0)\dagger}_1
-\hat{\bar{B}}^{(0)\dagger}_2)]\nonumber\\
&=&\cosh r\hat{B}^{(0)}_1+\sinh r \hat{B}^{(0)\dagger}_2,\nonumber\\
\hat{B}_2&=&\case{1}{\sqrt{2}}(\hat{\bar{B}}_1-\hat{\bar{B}}_2)\nonumber\\
&=&\case{1}{\sqrt{2}}[\cosh r(\hat{\bar{B}}^{(0)}_1-\hat{\bar{B}}^{(0)}_2)
+\sinh r(\hat{\bar{B}}^{(0)\dagger}_1
+\hat{\bar{B}}^{(0)\dagger}_2)]\nonumber\\
&=&\cosh r\hat{B}^{(0)}_2+\sinh r \hat{B}^{(0)\dagger}_1,
\end{eqnarray}
and
\begin{eqnarray}\label{1.22}
\hat{X}_1&=&\case{1}{\sqrt{2}}(\hat{\bar{X}}_1+\hat{\bar{X}}_2)=\case{1}
{\sqrt{2}}(e^{r} \hat{\bar{X}}^{(0)}_1+e^{-r} \hat{\bar{X}}^{(0)}_2),
\nonumber\\
\hat{P}_1&=&\case{1}{\sqrt{2}}(\hat{\bar{P}}_1+\hat{\bar{P}}_2)=\case{1}
{\sqrt{2}}(e^{-r} \hat{\bar{P}}^{(0)}_1+e^{r} \hat{\bar{P}}^{(0)}_2),
\nonumber\\
\hat{X}_2&=&\case{1}{\sqrt{2}}(\hat{\bar{X}}_1-\hat{\bar{X}}_2)=\case{1}
{\sqrt{2}}(e^{r} \hat{\bar{X}}^{(0)}_1-e^{-r} \hat{\bar{X}}^{(0)}_2),
\nonumber\\
\hat{P}_2&=&\case{1}{\sqrt{2}}(\hat{\bar{P}}_1-\hat{\bar{P}}_2)=\case{1}
{\sqrt{2}}(e^{-r} \hat{\bar{P}}^{(0)}_1-e^{r} \hat{\bar{P}}^{(0)}_2),
\end{eqnarray}
as the two-mode squeezed state in Eqs.~(\ref{1.2}). The coupled modes in 
Eqs.~(\ref{1.21}) expressed in terms of $\hat{B}^{(0)}_1$ and 
$\hat{B}^{(0)}_2$ are the two NOPA output modes of Eq.~(\ref{1.12}), 
if $\Omega\to 0$ and $G(0)=\cosh r$, $g(0)=\sinh r$. 

More generally, for $\Omega\neq 0$, the quadratures corresponding 
to Eqs.~(\ref{1.18}),
\begin{eqnarray}\label{1.23}
&&\hat{\bar{X}}_1(\Omega)=[G(\Omega)+g(\Omega)] \hat{\bar{X}}^{(0)}_1
(\Omega),\;\;\;\hat{\bar{P}}_1(\Omega)=[G(\Omega)-g(\Omega)]  
\hat{\bar{P}}^{(0)}_1(\Omega),
\nonumber\\
&&\hat{\bar{X}}_2(\Omega)=[G(\Omega)-g(\Omega)] \hat{\bar{X}}^{(0)}_2
(\Omega),\;\;\;\hat{\bar{P}}_2(\Omega)=[G(\Omega)+g(\Omega)]  
\hat{\bar{P}}^{(0)}_2(\Omega),
\end{eqnarray}
are coupled to yield
\begin{eqnarray}\label{1.24}
\hat{X}_1(\Omega)&=&\case{1}{\sqrt{2}}[G(\Omega)+g(\Omega)] 
\hat{\bar{X}}^{(0)}_1(\Omega)+\case{1}{\sqrt{2}}[G(\Omega)-g(\Omega)] 
\hat{\bar{X}}^{(0)}_2(\Omega),\nonumber\\
\hat{P}_1(\Omega)&=&\case{1}{\sqrt{2}}[G(\Omega)-g(\Omega)] 
\hat{\bar{P}}^{(0)}_1(\Omega)+\case{1}{\sqrt{2}}[G(\Omega)+g(\Omega)] 
\hat{\bar{P}}^{(0)}_2(\Omega),\nonumber\\
\hat{X}_2(\Omega)&=&\case{1}{\sqrt{2}}[G(\Omega)+g(\Omega)] 
\hat{\bar{X}}^{(0)}_1(\Omega)-\case{1}{\sqrt{2}}[G(\Omega)-g(\Omega)] 
\hat{\bar{X}}^{(0)}_2(\Omega),\nonumber\\
\hat{P}_2(\Omega)&=&\case{1}{\sqrt{2}}[G(\Omega)-g(\Omega)] 
\hat{\bar{P}}^{(0)}_1(\Omega)-\case{1}{\sqrt{2}}[G(\Omega)+g(\Omega)]
\hat{\bar{P}}^{(0)}_2(\Omega).
\end{eqnarray}
The quadratures in Eqs.~(\ref{1.24}) are precisely the NOPA output 
quadratures of Eqs.~(\ref{1.15}) as anticipated. With the functions 
$G(\Omega)$ and $g(\Omega)$ of Eqs.~(\ref{1.13}), we obtain
\begin{eqnarray}\label{1.25}
G(\Omega)-g(\Omega)&=&\frac{\gamma/2 - \kappa + i\Omega}{\gamma/2 + 
\kappa - i\Omega}\,\,\,,\nonumber\\
G(\Omega)+g(\Omega)&=&\frac{(\gamma/2 + \kappa)^2 + \Omega^2}
{(\gamma/2-i \Omega)^2-\kappa^2}\,\,\,.
\end{eqnarray}
For the limits $\Omega\to 0$, $\kappa\to\gamma/2$ (the limit of 
infinite squeezing), we obtain $[G(\Omega)-g(\Omega)]\to 0$ and 
$[G(\Omega)+g(\Omega)]\to\infty$. If $\Omega\to 0$, $\kappa\to 0$ 
(the classical limit of no squeezing), then $[G(\Omega)-g(\Omega)]\to 1$ 
and  $[G(\Omega)+g(\Omega)]\to 1$. Thus for $\Omega\to 0$, 
Eqs.~(\ref{1.24}) in the above mentioned limits correspond to 
Eqs.~(\ref{1.22}) in the analogous limits $r\to\infty$ 
(infinite squeezing) and $r\to 0$ (no squeezing). 
For large squeezing, apparently the individual modes of the 
``broadband two-mode squeezed state" in Eqs.~(\ref{1.24}) are very 
noisy. In general, the input vacuum modes are amplified in the NOPA, 
resulting in output modes with large fluctuations. But the correlations 
between the two modes increase simultaneously, so that $[\hat{X}_1
(\Omega)-\hat{X}_2(\Omega)]\to 0$ and $[\hat{P}_1(\Omega)+\hat{P}_2
(\Omega)]\to 0$ for $\Omega\to 0$ and $\kappa\to\gamma/2$.

The squeezing spectra of the independently squeezed modes can be derived
from Eqs.~(\ref{1.23}) and are given by the spectral variances
\begin{eqnarray}\label{spectra}
\langle\Delta\hat{\bar{X}}_1^{\dagger}(\Omega)\Delta\hat{\bar{X}}_1
(\Omega')\rangle&=&\langle\Delta\hat{\bar{P}}_2^{\dagger}(\Omega)
\Delta\hat{\bar{P}}_2(\Omega')\rangle=
\delta(\Omega-\Omega')|S_+(\Omega)|^2\langle\Delta\hat{X}^2
\rangle_{\rm vacuum},\nonumber\\
\langle\Delta\hat{\bar{X}}_2^{\dagger}(\Omega)\Delta\hat{\bar{X}}_2
(\Omega')\rangle&=&\langle\Delta\hat{\bar{P}}_1^{\dagger}(\Omega)
\Delta\hat{\bar{P}}_1(\Omega')\rangle=
\delta(\Omega-\Omega')|S_-(\Omega)|^2\langle\Delta\hat{X}^2
\rangle_{\rm vacuum},
\end{eqnarray}
here with $|S_+(\Omega)|^2=|G(\Omega)+g(\Omega)|^2$ and
$|S_-(\Omega)|^2=|G(\Omega)-g(\Omega)|^2$ ($\langle\Delta\hat{X}^2
\rangle_{\rm vacuum}=\case{1}{4}$).
In general, Eqs.~(\ref{spectra}) may define arbitrary squeezing spectra
of two statistically identical but independent broadband 
squeezed states. The two corresponding squeezed modes 
\begin{eqnarray}\label{general}
&&\hat{\bar{X}}_1(\Omega)=S_+(\Omega) \hat{\bar{X}}^{(0)}_1
(\Omega),\;\;\;\hat{\bar{P}}_1(\Omega)=S_-(\Omega) 
\hat{\bar{P}}^{(0)}_1(\Omega),
\nonumber\\
&&\hat{\bar{X}}_2(\Omega)=S_-(\Omega) \hat{\bar{X}}^{(0)}_2
(\Omega),\;\;\;\hat{\bar{P}}_2(\Omega)=S_+(\Omega) 
\hat{\bar{P}}^{(0)}_2(\Omega),
\end{eqnarray}
where $S_-(\Omega)$ refers to the quiet quadratures and $S_+(\Omega)$
to the noisy ones, can be used as EPR source for the following broadband 
teleportation scheme when they are combined at a beam splitter:
\begin{eqnarray}\label{general2}
\hat{X}_1(\Omega)&=&\case{1}{\sqrt{2}}S_+(\Omega) 
\hat{\bar{X}}^{(0)}_1(\Omega)+\case{1}{\sqrt{2}}S_-(\Omega) 
\hat{\bar{X}}^{(0)}_2(\Omega),\nonumber\\
\hat{P}_1(\Omega)&=&\case{1}{\sqrt{2}}S_-(\Omega) 
\hat{\bar{P}}^{(0)}_1(\Omega)+\case{1}{\sqrt{2}}S_+(\Omega) 
\hat{\bar{P}}^{(0)}_2(\Omega),\nonumber\\
\hat{X}_2(\Omega)&=&\case{1}{\sqrt{2}}S_+(\Omega)
\hat{\bar{X}}^{(0)}_1(\Omega)-\case{1}{\sqrt{2}}S_-(\Omega) 
\hat{\bar{X}}^{(0)}_2(\Omega),\nonumber\\
\hat{P}_2(\Omega)&=&\case{1}{\sqrt{2}}S_-(\Omega) 
\hat{\bar{P}}^{(0)}_1(\Omega)-\case{1}{\sqrt{2}}S_+(\Omega)
\hat{\bar{P}}^{(0)}_2(\Omega).
\end{eqnarray}
Before obtaining this ``broadband two-mode squeezed vacuum state,''
the squeezing of the two initial modes may be generated by
any nonlinear interaction, e.g., apart from the OPA, also
by four-wave mixing in a cavity \cite{Slush}.  

\section{Teleportation of a broadband field}

For the teleportation of an electromagnetic field with finite bandwidth, 
the EPR state shared by Alice and Bob should be a broadband two-mode 
squeezed state, as discussed in the previous section.
The incoming electromagnetic field 
to be teleported $\hat{E}_{\rm in}(z,t)=\hat{E}_{\rm in}^{(+)}(z,t)+
\hat{E}_{\rm in}^{(-)}(z,t)$, traveling in positive-z direction and 
having a single polarization, can be described by the positive-frequency 
part
\begin{eqnarray}\label{field}
\hat{E}_{\rm in}^{(+)}(z,t)&=&[\hat{E}_{\rm in}^{(-)}(z,t)]^{\dagger}=
\int_{\rm W} d\omega\frac{1}{\sqrt{2\pi}}\left(\frac{u\hbar\omega}
{2cA_{\rm tr}}\right)^{1/2}\hat{b}_{\rm in}(\omega)e^{-i\omega(t-z/c)}.
\end{eqnarray}
The integral runs over a relevant bandwidth W centered on $\omega_0$, 
$A_{\rm tr}$ represents the transverse structure of the field and $u$ is 
a units-dependent constant (in Gaussian units $u=4\pi$) \cite{Schum}.
The annihilation and creation operators $\hat{b}_{\rm in}(\omega)$ and
$\hat{b}_{\rm in}^{\dagger}(\omega)$ satisfy the commutation relations
$[\hat{b}_{\rm in}(\omega),\hat{b}_{\rm in}(\omega')]=0$ and
$[\hat{b}_{\rm in}(\omega),\hat{b}_{\rm in}^{\dagger}(\omega')]=\delta
(\omega-\omega')$. 
The incoming electromagnetic field may now be described in a rotating 
frame as
\begin{eqnarray}\label{1.26}
\hat{B}_{\rm in}(t)=\hat{X}_{\rm in}(t)+i\hat{P}_{\rm in}(t)=
[\hat{x}_{\rm in}(t)+i\hat{p}_{\rm in}(t)]e^{i\omega_0t}=
\hat{b}_{\rm in}(t)e^{i\omega_0t},
\end{eqnarray}
as in Eq.~(\ref{1.9}) with
\begin{eqnarray}\label{1.27}
\hat{B}_{\rm in}(\Omega)=\frac{1}{\sqrt{2\pi}}\int dt\hat{B}_{\rm in}(t)
e^{i\Omega t},
\end{eqnarray}
as in Eq.~(\ref{1.10}) and commutation relations $[\hat{B}_{\rm in}
(\Omega),\hat{B}_{\rm in}(\Omega')]=0$, $[\hat{B}_{\rm in}
(\Omega),\hat{B}_{\rm in}^{\dagger}(\Omega')]=\delta(\Omega - \Omega')$.

Of course, the unknown input field is not completely arbitrary.
In the case of an EPR state from the NOPA, we will see
that for successful quantum teleportation, the center of the input field's
spectral range W should be around the NOPA center frequency $\omega_0$
(half the pump frequency of the NOPA). Further, as we shall see, its 
spectral width should be small with respect to the NOPA bandwidth to 
benefit from the EPR correlations of the NOPA output. As for the
transverse structure and the single polarization of the input field, we 
assume that both are known to all participants.

In spite of these complications, the teleportation protocol is performed
in a fashion almost identical to the zero-bandwidth case. The EPR state 
of modes 1 and 2 is produced either directly as the NOPA output or by the 
superposition of two independently squeezed beams, as discussed in the
preceding section. Mode 1 is sent to Alice and mode 2 is sent to Bob 
(see Fig.~1) where for the case of the NOPA, these modes correspond to
two orthogonal polarizations. Alice arranges to superimpose mode 1 with 
the unknown input field at a 50/50 beam splitter, 
yielding for the relevant quadratures
\begin{eqnarray}\label{1.28}
\hat{X}_{\rm u}(\Omega)&=&\case{1}{\sqrt{2}}\hat{X}_{\rm in}(\Omega)
-\case{1}{\sqrt{2}}\hat{X}_1(\Omega),\nonumber\\
\hat{P}_{\rm v}(\Omega)&=&\case{1}{\sqrt{2}}\hat{P}_{\rm in}(\Omega)+
\case{1}{\sqrt{2}}\hat{P}_1(\Omega).
\end{eqnarray}
Using Eqs.~(\ref{1.28}) we will find it useful to write the quadrature 
operators of Bob's mode 2 as
\begin{eqnarray}\label{Bob2}
\hat{X}_2(\Omega)&=&\hat{X}_{\rm in}(\Omega)-
[\hat{X}_1(\Omega)-\hat{X}_2(\Omega)]-
\sqrt{2}\hat{X}_{\rm u}(\Omega)\nonumber\\
&=&\hat{X}_{\rm in}(\Omega)-
\sqrt{2}S_-(\Omega)\hat{\bar{X}}^{(0)}_2(\Omega)-
\sqrt{2}\hat{X}_{\rm u}(\Omega),\nonumber\\
\hat{P}_2(\Omega)&=&\hat{P}_{\rm in}(\Omega)+
[\hat{P}_1(\Omega)+\hat{P}_2(\Omega)]-
\sqrt{2}\hat{P}_{\rm v}(\Omega)\nonumber\\
&=&\hat{P}_{\rm in}(\Omega)+
\sqrt{2}S_-(\Omega)\hat{\bar{P}}^{(0)}_1(\Omega)-
\sqrt{2}\hat{P}_{\rm v}(\Omega).
\end{eqnarray}
Here we have used Eqs.~(\ref{general2}).
How is Alice's ``Bell detection'' which yields classical photocurrents
performed? The photocurrent operators for the two homodyne detections,
$\hat{i}_{\rm u}(t)\propto |E_{\rm LO}^{X}|\hat{X}_{\rm u}(t)$ and
$\hat{i}_{\rm v}(t)\propto |E_{\rm LO}^{P}|\hat{P}_{\rm v}(t)$,
can be written (without loss of generality we assume $\Omega>0$) as
\begin{eqnarray}\label{currents}
\hat{i}_{\rm u}(t)&\propto&|E_{\rm LO}^{X}|\int_{\rm W} d\Omega\,
h_{\rm el}(\Omega)\left[\hat{X}_{\rm u}(\Omega)e^{-i\Omega t}+
\hat{X}^{\dagger}_{\rm u}(\Omega)e^{i\Omega t}\right],\nonumber\\
\hat{i}_{\rm v}(t)&\propto&|E_{\rm LO}^{P}|\int_{\rm W} d\Omega\,
h_{\rm el}(\Omega)\left[\hat{P}_{\rm v}(\Omega)e^{-i\Omega t}+
\hat{P}^{\dagger}_{\rm v}(\Omega)e^{i\Omega t}\right],
\end{eqnarray} 
with a noiseless, classical local oscillator (LO) and $h_{\rm el}(\Omega)$
representing the detectors' responses within their electronic bandwidths
$\Delta\Omega_{\rm el}$: $h_{\rm el}(\Omega)=1$ for $\Omega\leq
\Delta\Omega_{\rm el}$ and zero otherwise. We assume that the relevant
bandwidth W ($\sim$ MHz) is fully covered by the electronic bandwidth
of the detectors ($\sim$ GHz). Therefore, $h_{\rm el}(\Omega)\equiv 1$
in Eqs.~(\ref{currents}).
Continuously in time, these photocurrents are measured and fed-forward
to Bob via a classical channel with sufficient RF bandwidth.
Each of them must be viewed as complex quantities in order to respect
the RF phase. 
The whole feedforward process, continuously performed in the time domain
(i.e., performed every inverse-bandwidth time),
includes Alice's detections, her classical transmission and corresponding
amplitude and phase modulations of Bob's EPR beam.
Any {\it relative} delays between the classical information conveyed by 
Alice and Bob's EPR beam must be such that $\Delta t\ll 1/\Delta\Omega$
with the inverse bandwidth of the EPR source $1/\Delta\Omega$
(for an EPR state from the NOPA: $\Delta t\ll\gamma^{-1}$).
Expressed in the frequency domain, the final modulations can be described
by the classical ``displacements''
\begin{eqnarray}\label{1.29}
\hat{X}_2(\Omega)\longrightarrow\hat{X}_{\rm tel}(\Omega)&=&\hat{X}_2
(\Omega)+\Gamma(\Omega)\sqrt{2}X_{\rm u}(\Omega),\nonumber\\
\hat{P}_2(\Omega)\longrightarrow\hat{P}_{\rm tel}(\Omega)&=&\hat{P}_2
(\Omega)+\Gamma(\Omega)\sqrt{2}P_{\rm v}(\Omega).
\end{eqnarray}
The parameter $\Gamma(\Omega)$ is again a suitably normalized gain
(now, in general, depending on $\Omega$).

For $\Gamma(\Omega)=1$, Bob's displacements from Eqs.~(\ref{1.29})
exactly eliminate $\hat{X}_{\rm u}(\Omega)$ and $\hat{P}_{\rm v}(\Omega)$ 
in Eqs.~(\ref{Bob2}). The same applies to the Hermitian conjugate
versions of Eqs.~(\ref{Bob2}) and Eqs.~(\ref{1.29}).
We obtain the teleported field
\begin{eqnarray}\label{1.30}
\hat{X}_{\rm tel}(\Omega)&=&\hat{X}_{\rm in}(\Omega)-
\sqrt{2}S_-(\Omega)\hat{\bar{X}}^{(0)}_2(\Omega),\nonumber\\
\hat{P}_{\rm tel}(\Omega)&=&\hat{P}_{\rm in}(\Omega)+
\sqrt{2}S_-(\Omega)\hat{\bar{P}}^{(0)}_1(\Omega).
\end{eqnarray}
For an arbitrary gain $\Gamma(\Omega)$, the teleported field becomes
\begin{eqnarray}\label{arbgain}
\hat{X}_{\rm tel}(\Omega)=\Gamma(\Omega)\hat{X}_{\rm in}(\Omega)
&-&\frac{\Gamma(\Omega)-1}{\sqrt{2}}S_+(\Omega) 
\hat{\bar{X}}^{(0)}_1(\Omega)\nonumber\\
&-&\frac{\Gamma(\Omega)+1}{\sqrt{2}}S_-(\Omega)
\hat{\bar{X}}^{(0)}_2(\Omega),\nonumber\\
\hat{P}_{\rm tel}(\Omega)=\Gamma(\Omega)\hat{P}_{\rm in}(\Omega)
&+&\frac{\Gamma(\Omega)-1}{\sqrt{2}}S_+(\Omega)
\hat{\bar{P}}^{(0)}_2(\Omega)\nonumber\\
&+&\frac{\Gamma(\Omega)+1}{\sqrt{2}}S_-(\Omega)
\hat{\bar{P}}^{(0)}_1(\Omega).
\end{eqnarray}
In general, these equations contain non-Hermitian operators with 
nonreal coefficients. Let us assume an EPR state from the NOPA,
$S_{\pm}(\Omega)=G(\Omega)\pm g(\Omega)$.
In the zero-bandwidth limit, the 
quadrature operators are Hermitian and the coefficients in
Eqs.~(\ref{1.30}) and Eqs.~(\ref{arbgain}) are real.
For $\Omega\to 0$ and $\Gamma(\Omega)=1$, 
the teleported quadratures computed from the above
equations are, in agreement with the zero-bandwidth results, given by 
$\hat{X}_{\rm tel}=\hat{X}_{\rm in}$ and
$\hat{P}_{\rm tel}=\hat{P}_{\rm in}$, if 
$\kappa\to\gamma/2$ and hence $[G(\Omega)-g(\Omega)]\to 0$ (infinite 
squeezing). Thus, for zero bandwidth and an infinite degree of EPR 
correlations, Alice's unknown quantum state of mode ``in'' is exactly 
reconstituted by Bob after generating the output mode ``tel'' through 
unit-gain displacements.
However, we are particularly interested in the physical case of 
finite bandwidth. Apparently, in unit-gain teleportation,
the complete disappearance of the two classical quduties 
for perfect teleportation requires $\Omega=0$ (with an EPR state from the 
NOPA). Does this mean an increasing bandwidth always leads to 
deteriorating quantum teleportation? In order to make quantitative 
statements about this issue, we consider input states with a 
coherent amplitude (unit-gain teleportation) and calculate 
the spectral variances of the teleported quadratures for a coherent-state
input to obtain a ``fidelity spectrum.'' 

\subsection{Teleporting broadband Gaussian fields with a coherent amplitude}

Let us employ teleportation equations for the real and imaginary parts 
of the non-Hermitian quadrature operators. In order to achieve a nonzero
average fidelity when teleporting fields with a coherent amplitude,
we assume $\Gamma(\Omega)=1$. According to 
Eqs.~(\ref{1.30}), the real and imaginary parts of the teleported
quadratures are
\begin{eqnarray}\label{1.31}
{\rm Re}\hat{X}_{\rm tel}(\Omega)=
{\rm Re}\hat{X}_{\rm in}(\Omega)
&-&\sqrt{2}{\rm Re}[S_-(\Omega)]{\rm Re}\hat{\bar{X}}^{(0)}_2(\Omega)
\nonumber\\
&+&\sqrt{2}{\rm Im}[S_-(\Omega)]{\rm Im}\hat{\bar{X}}^{(0)}_2(\Omega),
\nonumber\\
{\rm Re}\hat{P}_{\rm tel}(\Omega)=
{\rm Re}\hat{P}_{\rm in}(\Omega)
&+&\sqrt{2}{\rm Re}[S_-(\Omega)]{\rm Re}\hat{\bar{P}}^{(0)}_1(\Omega)
\nonumber\\
&-&\sqrt{2}{\rm Im}[S_-(\Omega)]{\rm Im}\hat{\bar{P}}^{(0)}_1(\Omega),
\nonumber\\
{\rm Im}\hat{X}_{\rm tel}(\Omega)=
{\rm Im}\hat{X}_{\rm in}(\Omega)
&-&\sqrt{2}{\rm Im}[S_-(\Omega)]{\rm Re}\hat{\bar{X}}^{(0)}_2(\Omega)
\nonumber\\
&-&\sqrt{2}{\rm Re}[S_-(\Omega)]{\rm Im}\hat{\bar{X}}^{(0)}_2(\Omega),
\nonumber\\
{\rm Im}\hat{P}_{\rm tel}(\Omega)=
{\rm Im}\hat{P}_{\rm in}(\Omega)
&+&\sqrt{2}{\rm Im}[S_-(\Omega)]{\rm Re}\hat{\bar{P}}^{(0)}_1(\Omega)
\nonumber\\
&+&\sqrt{2}{\rm Re}[S_-(\Omega)]{\rm Im}\hat{\bar{P}}^{(0)}_1(\Omega).
\end{eqnarray}
Their only nontrivial commutators are
\begin{eqnarray}\label{1.32}
[{\rm Re}\hat{X}_j(\Omega),{\rm Re}\hat{P}_j(\Omega')]=[{\rm Im}
\hat{X}_j(\Omega),{\rm Im}\hat{P}_j(\Omega')]=(i/4)\;\delta(\Omega-\Omega'),
\end{eqnarray}
where we have used Eqs.~(\ref{1.14}) and $[\hat{B}_j(\Omega),
\hat{B}_j^{\dagger}(\Omega')]=\delta(\Omega - \Omega')$.

We define spectral variances similar to Eq.~(\ref{telin}), 
\begin{eqnarray}\label{def1}
\frac{\langle\Delta[{\rm Re}\hat{X}_{\rm tel}(\Omega)-
{\rm Re}\hat{X}_{\rm in}(\Omega)]\Delta[{\rm Re}
\hat{X}_{\rm tel}(\Omega')-{\rm Re}\hat{X}_{\rm in}(\Omega')]
\rangle}{\langle\Delta{\rm Re}\hat{X}^2\rangle
_{\rm vacuum}}&\equiv&\delta(\Omega-\Omega')
V_{\rm tel,in}^{{\rm Re}\hat{X}}(\Omega).
\end{eqnarray}
We analogously define 
$V_{\rm tel,in}^{{\rm Re}\hat{P}}(\Omega)$, $V_{\rm tel,in}^{{\rm Im}
\hat{X}}(\Omega)$, and $V_{\rm tel,in}^{{\rm Im}\hat{P}}(\Omega)$ with 
${\rm Re}\hat{X}\rightarrow{\rm Re}\hat{P}$ etc. throughout. 

From Eqs.~(\ref{1.31}), we obtain
\begin{eqnarray}\label{1.34}
V_{\rm tel,in}^{{\rm Re}\hat{X}}(\Omega)&=&
V_{\rm tel,in}^{{\rm Re}\hat{P}}(\Omega)=
V_{\rm tel,in}^{{\rm Im}\hat{X}}(\Omega)=
V_{\rm tel,in}^{{\rm Im}\hat{P}}(\Omega)\nonumber\\
&=&2\;|S_-(\Omega)|^2.
\end{eqnarray}
Here we have used that
\begin{eqnarray}
\langle\Delta{\rm Re}\hat{\bar{X}}^{(0)}_j(\Omega)
\Delta{\rm Re}\hat{\bar{X}}^{(0)}_j(\Omega')\rangle&=&
\delta(\Omega-\Omega')\langle\Delta{\rm Re}\hat{X}^2\rangle
_{\rm vacuum}=\nonumber\\
\langle\Delta{\rm Im}\hat{\bar{X}}^{(0)}_j(\Omega)
\Delta{\rm Im}\hat{\bar{X}}^{(0)}_j(\Omega')\rangle&=&
\delta(\Omega-\Omega')\langle\Delta{\rm Im}\hat{X}^2\rangle
_{\rm vacuum}, 
\end{eqnarray}
and analogously for the other quadrature, and
\begin{eqnarray}
\langle\Delta{\rm Re}\hat{\bar{X}}^{(0)}_j(\Omega)
\Delta{\rm Im}\hat{\bar{X}}^{(0)}_j(\Omega')\rangle=
\langle\Delta{\rm Re}\hat{\bar{P}}^{(0)}_j(\Omega)
\Delta{\rm Im}\hat{\bar{P}}^{(0)}_j(\Omega')\rangle=0\,\,\,.
\end{eqnarray}
Thus, for unit-gain teleportation at all frequencies, it turns out that 
{\it the variance of each teleported quadrature is given 
by the variance of the input quadrature plus 
twice the squeezing spectrum of the quiet quadrature of a decoupled 
mode in a ``broadband squeezed state''} as in Eqs.~(\ref{general}).
The excess noise in each teleported quadrature after the teleportation 
process is, relative to the vacuum noise, {\it twice} the 
squeezing spectrum $|S_-(\Omega)|^2$ from Eqs.~(\ref{spectra}).

We also obtain these results by directly defining
\begin{eqnarray}\label{def2}
\frac{\langle\Delta[\hat{X}^{\dagger}_{\rm tel}(\Omega)-
\hat{X}^{\dagger}_{\rm in}(\Omega)]\Delta[\hat{X}_{\rm tel}(\Omega')-
\hat{X}_{\rm in}(\Omega')]\rangle}{\langle\Delta\hat{X}^2
\rangle_{\rm vacuum}}&\equiv&\delta(\Omega-\Omega')
V_{\rm tel,in}^{\hat{X}}(\Omega).
\end{eqnarray}
We analogously define
$V_{\rm tel,in}^{\hat{P}}(\Omega)$ with $\hat{X}\rightarrow\hat{P}$ 
throughout. Using Eqs.~(\ref{1.30}), these variances become for 
$\Gamma(\Omega)=1$
\begin{eqnarray}\label{quadrvar}
V_{\rm tel,in}^{\hat{X}}(\Omega)&=&
V_{\rm tel,in}^{\hat{P}}(\Omega)=2\;|S_-(\Omega)|^2.
\end{eqnarray}
We calculate some limits for $V_{\rm tel,in}^{\hat{X}}(\Omega)$ of 
Eq.~(\ref{quadrvar}), assuming an EPR state from the NOPA,
$S_-(\Omega)=G(\Omega)-g(\Omega)$.
Since $V_{\rm tel,in}^{\hat{X}}(\Omega)=
V_{\rm tel,in}^{\hat{P}}(\Omega)$ and $\Gamma(\Omega)=1$, we can name
the limits according to the criterion of Eq.~(\ref{limit}).\\
{\bf Classical teleportation, $\kappa\to 0$}:\\
$V_{\rm tel,in}^{\hat{X}}(\Omega)=2$, which is independent of the 
modulation frequency $\Omega$.\\
{\bf Zero-bandwidth quantum teleportation, $\Omega\to 0$, $\kappa> 0$}:\\
$V_{\rm tel,in}^{\hat{X}}(\Omega)=2\,
[1-2\kappa\gamma/(\kappa+\gamma/2)^2]$, and in the 
ideal case of infinite squeezing $\kappa\to \gamma/2$: $V_{\rm tel,in}
^{\hat{X}}(\Omega)=0$.\\
{\bf Broadband quantum teleportation, $\Omega> 0$, $\kappa> 0$}:\\
$V_{\rm tel,in}^{\hat{X}}(\Omega)=2\,
\{1-2\kappa\gamma/[(\kappa+\gamma/2)^2+\Omega^2]\}$, 
and in the ideal case $\kappa\to \gamma/2$: 
$V_{\rm tel,in}^{\hat{X}}(\Omega)=2\,[\Omega^2/(\gamma^2+\Omega^2)]$. 
So it turns out that also for finite bandwidth ideal quantum teleportation 
can be approached provided $\Omega\ll \gamma$.

We can express $V_{\rm tel,in}^{\hat{X}}(\Omega)$ in terms of 
experimental parameters relevant to the NOPA. For this purpose, 
we use the dimensionless quantities from Ref.~\onlinecite{Ou},
\begin{eqnarray}\label{1.35}
\epsilon=\frac{2\kappa}{\gamma+\rho}=
\sqrt{\frac{P_{\rm pump}}{P_{\rm thres}}}\,\,\,,\;\;\;\;
\omega=\frac{2\Omega}{\gamma+\rho}=\frac{\Omega}{2\pi}\,
\frac{2F_{\rm cav}}{\nu_{\rm FSR}}\,\,\,.
\end{eqnarray}
Here, $P_{\rm pump}$ is the pump power, $P_{\rm thres}$ is the threshold 
value, $F_{\rm cav}$ is the measured finesse of the cavity, 
$\nu_{\rm FSR}$ is its free spectral range, and the parameter $\rho$ 
describes cavity losses (see Fig.~4). Note that we now use $\omega$
as a normalized modulation frequency in contrast to Eq.~(\ref{field})
and the following commutators where it was the frequency of the field 
operators in the nonrotating frame.    

The spectral variances for the lossless case ($\rho=0$) 
can be written as a function of $\epsilon$ and $\omega$, namely,
\begin{eqnarray}\label{1.36}
V_{\rm tel,in}^{\hat{X}}(\epsilon,\omega)=
V_{\rm tel,in}^{\hat{P}}(\epsilon,\omega)=2\;\left[1-
\frac{4\epsilon}{(\epsilon+1)^2 + \omega^2}\right].
\end{eqnarray}
Now, the classical limit is $\epsilon\to 0$ ($V_{\rm tel,in}
^{\hat{X}}=2$, independent of $\omega$) and the ideal case
is $\epsilon\to 1$ [$V_{\rm tel,in}^{\hat{X}}(\epsilon,\omega)=
2\omega^2/(4+\omega^2)$]. Obviously, 
perfect quantum teleportation is achieved for $\epsilon\to 1$ and 
$\omega\to 0$. In fact, this limit can also be approached for finite 
$\Omega\neq 0$ provided $\omega\ll 1$ or $\Omega\ll \gamma$. Note 
that this condition is not specific to broadband teleportation, but is 
simply the condition for broadband squeezing, i.e., for the generation of 
highly squeezed quadratures at nonzero modulation frequencies $\Omega$.

Let us now assume coherent-state inputs
$\langle\Delta\hat{X}_{\rm in}^{\dagger}(\Omega)
\Delta\hat{X}_{\rm in}(\Omega')\rangle=
\langle\Delta\hat{P}_{\rm in}^{\dagger}(\Omega)
\Delta\hat{P}_{\rm in}(\Omega')\rangle=\case{1}{4}\delta(\Omega-\Omega')$
[and $\langle\Delta{\rm Re}\hat{X}_{\rm in}(\Omega)
\Delta{\rm Re}\hat{X}_{\rm in}(\Omega')\rangle=\case{1}{8}
\delta(\Omega-\Omega')$, etc.],
at all frequencies $\Omega$ in the relevant bandwidth W.
In order to obtain a spectrum of the fidelities in Eq.~(\ref{fid4})
with $\Gamma\rightarrow\Gamma(\Omega)=1$, we need the spectrum
of the Q functions of the teleported field with the spectral variances
$\sigma_x(\Omega)=\sigma_p(\Omega)=\case{1}{2}+\case{1}{4}
V_{\rm tel,in}^{\hat{X}}(\Omega)$.
We obtain the ``fidelity spectrum''
\begin{eqnarray}\label{fspec}
F(\Omega)=\frac{1}{1+|S_-(\Omega)|^2}\,\,\,.
\end{eqnarray}
Finally, with the new quantities
$\epsilon$ and $\omega$, the fidelity spectrum for quantum teleportation
of arbitrary broadband coherent states using broadband entanglement from the 
NOPA ($\rho=0$) is given by
\begin{eqnarray}\label{fidspec}
F(\epsilon,\omega)=\left[2-\frac{4\epsilon}{(\epsilon+1)^2 
+\omega^2}\right]^{-1}\,\,\,.
\end{eqnarray}
For different $\epsilon$ values, the spectrum of fidelities is shown
in Fig.~5. From the single-mode protocol (with ideal
detectors), we know that any nonzero
squeezing enables quantum teleportation and coherent-state inputs
can be teleported with $F=F_{\rm av}>\case{1}{2}$ for any $r>0$.
Correspondingly, the fidelity from Eq.~(\ref{fidspec}) exceeds 
$\case{1}{2}$ for any nonzero $\epsilon$ at all finite frequencies,
as, provided $\epsilon>0$, there is no squeezing at all only when
$\omega\to\infty$. However, we had assumed [see after Eqs.~(\ref{1.14}):
$\Omega\ll \omega_0$] modulation frequencies $\Omega$ much smaller than 
the NOPA center frequency $\omega_0$. In fact, for $\Omega\to
\omega_0$, squeezing becomes impossible at the frequency $\Omega$
\cite{Schum}. But also within the region $\Omega\ll \omega_0$,
effectively, the squeezing bandwith is
limited and hence as well the bandwith of quantum teleportation 
$\Delta\omega\equiv 2\omega_{\rm max}$ where $F(\omega)\approx\case{1}{2}$
($<0.51$)
for all $\omega>\omega_{\rm max}$ and $F(\omega)>\case{1}{2}$
($\geq 0.51$)
for all $\omega\leq\omega_{\rm max}$.
According to Fig.~5, we could say that the ``effective teleportation
bandwidth'' is just about $\Delta\omega\approx 5.8$ ($\epsilon=0.1$),
$\Delta\omega\approx 8.6$ ($\epsilon=0.2$),
$\Delta\omega\approx 12.4$ ($\epsilon=0.4$),
$\Delta\omega\approx 15.2$ ($\epsilon=0.6$), and
$\Delta\omega\approx 19.6$ ($\epsilon=1$). The maximum fidelities at
frequency $\omega=0$ are $F_{\rm max}\approx 0.6$ ($\epsilon=0.1$),
$F_{\rm max}\approx 0.69$ ($\epsilon=0.2$),
$F_{\rm max}\approx 0.84$ ($\epsilon=0.4$),
$F_{\rm max}\approx 0.94$ ($\epsilon=0.6$), and, of course,
$F_{\rm max}=1$ ($\epsilon=1$).
   
\subsection{Broadband entanglement swapping}

As discussed in Sec. III, we particularly want our teleportation device
to be capable of teleporting entanglement. We will present now the
broadband theory of this entanglement swapping for continuous variables,
as it was proposed in Ref.~\onlinecite{PvL} for single modes.
Before any detections (see Fig.~3), Alice (mode 1) and 
Claire (mode 2) share the
broadband two-mode squeezed state from Eqs.~(\ref{general2}), whereas
Claire (mode 3) and Bob (mode 4) share the corresponding entangled state
of modes 3 and 4 given by
\begin{eqnarray}\label{general3}
\hat{X}_3(\Omega)&=&\case{1}{\sqrt{2}}S_+(\Omega) 
\hat{\bar{X}}^{(0)}_3(\Omega)+\case{1}{\sqrt{2}}S_-(\Omega) 
\hat{\bar{X}}^{(0)}_4(\Omega),\nonumber\\
\hat{P}_3(\Omega)&=&\case{1}{\sqrt{2}}S_-(\Omega) 
\hat{\bar{P}}^{(0)}_3(\Omega)+\case{1}{\sqrt{2}}S_+(\Omega) 
\hat{\bar{P}}^{(0)}_4(\Omega),\nonumber\\
\hat{X}_4(\Omega)&=&\case{1}{\sqrt{2}}S_+(\Omega)
\hat{\bar{X}}^{(0)}_3(\Omega)-\case{1}{\sqrt{2}}S_-(\Omega) 
\hat{\bar{X}}^{(0)}_4(\Omega),\nonumber\\
\hat{P}_4(\Omega)&=&\case{1}{\sqrt{2}}S_-(\Omega) 
\hat{\bar{P}}^{(0)}_3(\Omega)-\case{1}{\sqrt{2}}S_+(\Omega)
\hat{\bar{P}}^{(0)}_4(\Omega).
\end{eqnarray}
Let us interpret the entanglement swapping here as quantum teleportation
of mode 2 to mode 4 using the entanglement of modes 3 and 4. This means
we want Bob to perform ``displacements'' based on the classical results
of Claire's Bell detection, i.e., the classical determination of
$\hat{X}_{\rm u}(\Omega)=[\hat{X}_2(\Omega)-\hat{X}_3(\Omega)]/\sqrt{2},
\hat{P}_{\rm v}(\Omega)=[\hat{P}_2(\Omega)+\hat{P}_3(\Omega)]/\sqrt{2}$.
These final ``displacements'' (amplitude and phase modulations) of mode 4
are crucial in order to reveal the entanglement from entanglement 
swapping and, for verification, to finally exploit it in a second round of
quantum teleportation using the previously unentangled modes 1 and 4
\cite{PvL}.
The entire teleportation process with arbitrary gain $\Gamma(\Omega)$
that led to Eqs.~(\ref{arbgain}), yields now, for the teleportation of 
mode 2 to mode 4, the teleported mode $4'$
[where in Eqs.~(\ref{arbgain}) simply 
$\hat{X}_{\rm tel}(\Omega)\rightarrow\hat{X}_4'(\Omega),
\hat{P}_{\rm tel}(\Omega)\rightarrow\hat{P}_4'(\Omega),
\hat{X}_{\rm in}(\Omega)\rightarrow\hat{X}_2(\Omega),
\hat{P}_{\rm in}(\Omega)\rightarrow\hat{P}_2(\Omega),
\hat{\bar{X}}^{(0)}_1(\Omega)\rightarrow\hat{\bar{X}}^{(0)}_3(\Omega),
\hat{\bar{P}}^{(0)}_1(\Omega)\rightarrow\hat{\bar{P}}^{(0)}_3(\Omega),
\hat{\bar{X}}^{(0)}_2(\Omega)\rightarrow\hat{\bar{X}}^{(0)}_4(\Omega),
\hat{\bar{P}}^{(0)}_2(\Omega)\rightarrow\hat{\bar{P}}^{(0)}_4(\Omega)$,
and $\Gamma(\Omega)\rightarrow\Gamma_{\rm swap}(\Omega)$],
\begin{eqnarray}\label{arbgain2}
\hat{X}_4'(\Omega)&=&\frac{\Gamma_{\rm swap}(\Omega)}{\sqrt{2}}
[S_+(\Omega)\hat{\bar{X}}^{(0)}_1(\Omega)-S_-(\Omega) 
\hat{\bar{X}}^{(0)}_2(\Omega)]\nonumber\\
&-&\frac{\Gamma_{\rm swap}(\Omega)-1}{\sqrt{2}}S_+(\Omega) 
\hat{\bar{X}}^{(0)}_3(\Omega)
-\frac{\Gamma_{\rm swap}(\Omega)+1}{\sqrt{2}}S_-(\Omega)
\hat{\bar{X}}^{(0)}_4(\Omega),\nonumber\\
\hat{P}_4'(\Omega)&=&\frac{\Gamma_{\rm swap}(\Omega)}{\sqrt{2}}
[S_-(\Omega) \hat{\bar{P}}^{(0)}_1(\Omega)-S_+(\Omega)
\hat{\bar{P}}^{(0)}_2(\Omega)]\nonumber\\
&+&\frac{\Gamma_{\rm swap}(\Omega)+1}{\sqrt{2}}S_-(\Omega)
\hat{\bar{P}}^{(0)}_3(\Omega)
+\frac{\Gamma_{\rm swap}(\Omega)-1}{\sqrt{2}}S_+(\Omega)
\hat{\bar{P}}^{(0)}_4(\Omega).
\end{eqnarray}
Provided entanglement swapping is successful, Alice and Bob can
use their modes 1 and $4'$ for a further quantum teleportation.
Assuming unit gain in this ``second teleportation,'' where the unknown
input state $\hat{X}_{\rm in}(\Omega)$, $\hat{P}_{\rm in}(\Omega)$
is to be teleported, the teleported field becomes
\begin{eqnarray}\label{arbgain3}
\hat{X}_{\rm tel}(\Omega)&=&\hat{X}_{\rm in}(\Omega)
+\frac{\Gamma_{\rm swap}(\Omega)-1}{\sqrt{2}}
S_+(\Omega)\hat{\bar{X}}^{(0)}_1(\Omega)
-\frac{\Gamma_{\rm swap}(\Omega)+1}{\sqrt{2}}
S_-(\Omega)\hat{\bar{X}}^{(0)}_2(\Omega)\nonumber\\
&-&\frac{\Gamma_{\rm swap}(\Omega)-1}{\sqrt{2}}S_+(\Omega) 
\hat{\bar{X}}^{(0)}_3(\Omega)
-\frac{\Gamma_{\rm swap}(\Omega)+1}{\sqrt{2}}S_-(\Omega)
\hat{\bar{X}}^{(0)}_4(\Omega),\nonumber\\
\hat{P}_{\rm tel}(\Omega)&=&\hat{P}_{\rm in}(\Omega)
+\frac{\Gamma_{\rm swap}(\Omega)+1}{\sqrt{2}}
S_-(\Omega)\hat{\bar{P}}^{(0)}_1(\Omega)
-\frac{\Gamma_{\rm swap}(\Omega)-1}{\sqrt{2}}
S_+(\Omega)\hat{\bar{P}}^{(0)}_2(\Omega)\nonumber\\
&+&\frac{\Gamma_{\rm swap}(\Omega)+1}{\sqrt{2}}S_-(\Omega)
\hat{\bar{P}}^{(0)}_3(\Omega)
+\frac{\Gamma_{\rm swap}(\Omega)-1}{\sqrt{2}}S_+(\Omega)
\hat{\bar{P}}^{(0)}_4(\Omega).
\end{eqnarray}
We calculate a fidelity spectrum for coherent-state inputs and obtain
\begin{eqnarray}\label{fidswap}
F(\Omega)=\{1&+&[\Gamma_{\rm swap}(\Omega)-1]^2|S_+(\Omega)|^2/2\nonumber\\
&+&[\Gamma_{\rm swap}(\Omega)+1]^2|S_-(\Omega)|^2/2\}^{-1}\,\,\,.
\end{eqnarray}
The optimum gain, depending on the amount of squeezing, that maximizes
this fidelity \cite{PvL} at different frequencies turns out to be
\begin{eqnarray}\label{gswap}
\Gamma_{\rm swap}(\Omega)=\frac{|S_+(\Omega)|^2-|S_-(\Omega)|^2}
{|S_+(\Omega)|^2+|S_-(\Omega)|^2}\,\,\,.
\end{eqnarray}
Let us now assume that the broadband entanglement comes from the NOPA
(two NOPA's with equal squeezing spectra),
$|S_-(\Omega)|^2\rightarrow |S_-(\epsilon,\omega)|^2=
1-4\epsilon/[(\epsilon+1)^2+\omega^2]$,
$|S_+(\Omega)|^2\rightarrow |S_+(\epsilon,\omega)|^2=
1+4\epsilon/[(\epsilon-1)^2+\omega^2]$.
The optimized fidelity then becomes
\begin{eqnarray}\label{optfidswap}
F_{\rm opt}(\epsilon,\omega)=
\left\{1+2\;\frac{[(\epsilon+1)^2+\omega^2][(\epsilon-1)^2+\omega^2]}
{[(\epsilon+1)^2+\omega^2]^2+[(\epsilon-1)^2+\omega^2]^2}\right\}^{-1}
\,\,\,.
\end{eqnarray}
The spectrum of these optimized fidelities is shown
in Fig.~6 for different $\epsilon$ values. Again, we know from
the single-mode protocol \cite{PvL} with ideal detectors that any nonzero
squeezing in both initial entanglement sources is sufficient for 
entanglement swapping to occur. In this case, mode 1 and $4'$ 
enable quantum teleportation and coherent-state inputs
can be teleported with $F=F_{\rm av}>\case{1}{2}$.
The fidelity from Eq.~(\ref{optfidswap}) is $\case{1}{2}$ for
$\epsilon=0$ and becomes $F_{\rm opt}(\epsilon,\omega)>\case{1}{2}$
for any $\epsilon>0$, provided that $\omega$ does not become infinite
(however, we had assumed $\Omega\ll \omega_0$).
In this sense, the squeezing or entanglement bandwidth is preserved
through entanglement swapping. At each frequency where the initial states
were squeezed and entangled, also the output state of modes 1 and $4'$
is entangled, but with less squeezing and worse quality of entanglement
(unless we had infinite squeezing in the initial states so that the
entanglement is perfectly teleported) \cite{PvL2}. 
Correspondingly, at frequencies with initially very small entanglement,
the entanglement becomes even smaller after entanglement swapping 
(but never vanishes completely). Thus, the effective bandwidth of
squeezing or entanglement decreases through entanglement swapping.
Then, compared to the teleportation bandwidth using broadband two-mode
squeezed states without entanglement swapping,
the bandwidth of teleportation using the output of entanglement
swapping is effectively smaller. 
The spectrum of the fidelities from Eq.~(\ref{optfidswap}) is narrower
and the ``effective teleportation bandwidth'' 
is now about $\Delta\omega\approx 1.2$ ($\epsilon=0.1$),
$\Delta\omega\approx 2.6$ ($\epsilon=0.2$),
$\Delta\omega\approx 4.2$ ($\epsilon=0.4$),
$\Delta\omega\approx 5.2$ ($\epsilon=0.6$),
and $\Delta\omega\approx 6.8$ ($\epsilon=1$). The maximum fidelities at
frequency $\omega=0$ are $F_{\rm max}\approx 0.52$ ($\epsilon=0.1$),
$F_{\rm max}\approx 0.57$ ($\epsilon=0.2$),
$F_{\rm max}\approx 0.74$ ($\epsilon=0.4$),
$F_{\rm max}\approx 0.89$ ($\epsilon=0.6$), and, still,
$F_{\rm max}=1$ ($\epsilon=1$).
 
\section{Cavity losses and Bell detector inefficiencies}

We extend the previous calculations and include losses for the particular
case of the NOPA cavity and inefficiencies in Alice's Bell detection. 
For this purpose, we use Eq.~(\ref{1.11}) for the outgoing NOPA modes.
We consider losses and inefficiencies for unit-gain teleportation
(teleportation of Gaussian states with a coherent amplitude).
For the case of entanglement swapping (nonunit-gain teleportation),
detector inefficiencies have been included in the single-mode
treatment of Ref.~\onlinecite{PvL}.
By superimposing the unknown input mode with the NOPA 
mode 1, the relevant quadratures from Eqs.~(\ref{1.28}) now become  
\begin{eqnarray}\label{1.40}
\hat{X}_{\rm u}(\Omega)&=&\frac{\eta}{\sqrt{2}}\hat{X}_{\rm in}(\Omega)
-\frac{\eta}{\sqrt{2}}\hat{X}_1(\Omega)+\sqrt{\frac{1-\eta^2}{2}}
\hat{X}^{(0)}_{D}(\Omega)+\sqrt{\frac{1-\eta^2}{2}}\hat{X}^{(0)}_{E}
(\Omega),\nonumber\\
\hat{P}_{\rm v}(\Omega)&=&\frac{\eta}{\sqrt{2}}\hat{P}_{\rm in}(\Omega)
+\frac{\eta}{\sqrt{2}}\hat{P}_1(\Omega)+\sqrt{\frac{1-\eta^2}{2}}
\hat{P}^{(0)}_{F}(\Omega)+\sqrt{\frac{1-\eta^2}{2}}\hat{P}^{(0)}_{G}
(\Omega).
\end{eqnarray}
The last two terms in each quadrature in Eqs.~(\ref{1.40}) represent 
additional vacua due to homodyne detection inefficiencies (the detector 
amplitude efficiency $\eta$ is assumed to be constant over the bandwidth 
of interest). 
Using Eqs.~(\ref{1.40}) it is useful to write the quadratures of NOPA 
mode 2 corresponding to Eq.~(\ref{1.11}) as
\begin{eqnarray}\label{mod2ineff}
\hat{X}_2(\Omega)&=&\hat{X}_{\rm in}(\Omega)-[G(\Omega)-
g(\Omega)] [\hat{X}^{(0)}_1(\Omega)-\hat{X}^{(0)}_2(\Omega)]\nonumber\\
&-&[\bar{G}(\Omega)-\bar{g}(\Omega)] [\hat{X}^{(0)}_{C,1}(\Omega)-
\hat{X}^{(0)}_{C,2}(\Omega)]+\sqrt{\frac{1-\eta^2}{\eta^2}}\hat{X}^{(0)}_{D}
(\Omega)+\sqrt{\frac{1-\eta^2}{\eta^2}}\hat{X}^{(0)}_{E}(\Omega)\nonumber\\
&-&\frac{\sqrt{2}}{\eta}\hat{X}_{\rm u}(\Omega),\nonumber\\
\hat{P}_2(\Omega)&=&\hat{P}_{\rm in}(\Omega)+[G(\Omega)-g(\Omega)]
[\hat{P}^{(0)}_1(\Omega)+\hat{P}^{(0)}_2(\Omega)]\nonumber\\
&+&[\bar{G}(\Omega)-\bar{g}(\Omega)] [\hat{P}^{(0)}_{C,1}(\Omega)+
\hat{P}^{(0)}_{C,2}(\Omega)]+\sqrt{
\frac{1-\eta^2}{\eta^2}}\hat{P}^{(0)}_{F}(\Omega)+\sqrt{\frac{1-
\eta^2}{\eta^2}}\hat{P}^{(0)}_{G}(\Omega)\nonumber\\
&-&\frac{\sqrt{2}}{\eta}\hat{P}_{\rm v}(\Omega),
\end{eqnarray}
where now \cite{Ou}
\begin{eqnarray}\label{1.41}
G(\Omega)&=&\frac{\displaystyle\kappa^2+\left({\gamma-\rho\over2}+i\Omega
\right)\left(\frac{\gamma+\rho}{2}-i\Omega\right)}{\displaystyle\left
(\frac{\gamma+\rho}{2}-i\Omega\right)^2-\kappa^2}\,\,\,,\nonumber\\
g(\Omega)&=&\frac{\kappa\gamma}{\displaystyle\left(\frac{\gamma+\rho}{2}-
i\Omega\right)^2-\kappa^2}\,\,\,,\nonumber\\
\bar{G}(\Omega)&=&\frac{\displaystyle\sqrt{\gamma\rho}\left(\frac
{\gamma+\rho}{2}
-i\Omega\right)}{\displaystyle\left(\frac{\gamma+\rho}{2}-i\Omega\right)^2
-\kappa^2}\,\,\,,\nonumber\\
\bar{g}(\Omega)&=&\frac{\kappa\sqrt{\gamma\rho}}{\displaystyle\left
(\frac{\gamma+\rho}{2}-i\Omega\right)^2-\kappa^2}\,\,\,,
\end{eqnarray}
still with $G(\Omega)=G^*(-\Omega)$, $g(\Omega)=g^*(-\Omega)$, and also 
$\bar{G}(\Omega)=\bar{G}^*(-\Omega)$, $\bar{g}(\Omega)=\bar{g}^*(-\Omega)$.
The quadratures $\hat{X}^{(0)}_{C,j}(\Omega)$ and 
$\hat{P}^{(0)}_{C,j}(\Omega)$ are those of the vacuum modes 
$\hat{C}^{(0)}_j(\Omega)$ in Eq.~(\ref{1.11}) according to 
Eqs.~(\ref{1.14}).

Again, $\hat{X}_{\rm u}(\Omega)$ and $\hat{P}_{\rm v}(\Omega)$ in 
Eqs.~(\ref{mod2ineff}) can be considered as classically determined
quantities $X_{\rm u}(\Omega)$ and $P_{\rm v}(\Omega)$ due to Alice's
measurements. The appropriate amplitude and 
phase modulations of mode 2 by Bob depending on the classical results of 
Alice's detections are described by
\begin{eqnarray}\label{1.42}
\hat{X}_2(\Omega)\longrightarrow\hat{X}_{\rm tel}(\Omega)&=&\hat{X}_2
(\Omega)+\Gamma(\Omega)\frac{\sqrt{2}}{\eta}X_{\rm u}(\Omega),\nonumber\\
\hat{P}_2(\Omega)\longrightarrow\hat{P}_{\rm tel}(\Omega)&=&\hat{P}_2
(\Omega)+\Gamma(\Omega)\frac{\sqrt{2}}{\eta}P_{\rm v}(\Omega).
\end{eqnarray}
For $\Gamma(\Omega)=1$, the teleported quadratures become
\begin{eqnarray}\label{1.43}
\hat{X}_{\rm tel}(\Omega)&=&\hat{X}_{\rm in}(\Omega)-[G(\Omega)-
g(\Omega)] [\hat{X}^{(0)}_1(\Omega)-\hat{X}^{(0)}_2(\Omega)]\nonumber\\
&-&[\bar{G}(\Omega)-\bar{g}(\Omega)] [\hat{X}^{(0)}_{C,1}(\Omega)-
\hat{X}^{(0)}_{C,2}(\Omega)]+\sqrt{\frac{1-\eta^2}{\eta^2}}\hat{X}^{(0)}_{D}
(\Omega)+\sqrt{\frac{1-\eta^2}{\eta^2}}\hat{X}^{(0)}_{E}(\Omega),\nonumber\\
\hat{P}_{\rm tel}(\Omega)&=&\hat{P}_{\rm in}(\Omega)+[G(\Omega)-g(\Omega)]
[\hat{P}^{(0)}_1(\Omega)+\hat{P}^{(0)}_2(\Omega)]\nonumber\\
&+&[\bar{G}(\Omega)-\bar{g}(\Omega)]
[\hat{P}^{(0)}_{C,1}(\Omega)+\hat{P}^{(0)}_{C,2}(\Omega)]+\sqrt{
\frac{1-\eta^2}{\eta^2}}\hat{P}^{(0)}_{F}(\Omega)+\sqrt{\frac{1-
\eta^2}{\eta^2}}\hat{P}^{(0)}_{G}(\Omega).
\end{eqnarray}

We calculate again spectral variances and obtain with the 
dimensionless variables of Eqs.~(\ref{1.35})
\begin{eqnarray}\label{1.44}
V_{\rm tel,in}^{\hat{X}}(\epsilon,\omega)=
V_{\rm tel,in}^{\hat{P}}(\epsilon,\omega)=
2\;\left[1-\frac{4\epsilon\beta}
{(\epsilon+1)^2 + \omega^2}\right]+2\,\frac{1-\eta^2}{\eta^2}\,\,\,,
\end{eqnarray}
where $\beta=\gamma/(\gamma+\rho)$ is a ``cavity escape efficiency'' 
which contains losses \cite{Ou}. With the spectral Q-function variances
of the teleported field $\sigma_x(\Omega)=\sigma_p(\Omega)=\case{1}{2}
+\case{1}{4}V_{\rm tel,in}^{\hat{X}}(\Omega)$, now for coherent-state
inputs, we find the fidelity spectrum (unit gain)
\begin{eqnarray}\label{fidspec2}
F(\epsilon,\omega)=\left[2-\frac{4\epsilon\beta}{(\epsilon+1)^2 
+\omega^2}+\frac{1-\eta^2}{\eta^2}\right]^{-1}\,\,\,.
\end{eqnarray}
Using the values $\epsilon=0.77$, $\omega=0.56$, and $\beta=0.9$, the 
measured values in the EPR experiment of Ref.~\onlinecite{Ou} 
for maximum pump power (but still below threshold), 
and a Bell detector efficiency $\eta^2=0.97$
(as in the teleportation experiment of Ref.~\onlinecite{Furu}), we obtain
$V_{\rm tel,in}^{\hat{X}}=V_{\rm tel,in}^{\hat{P}}=0.453$
and a fidelity $F=0.815$.
The measured value for the ``normalized analysis frequency'' $\omega=0.56$ 
corresponds to the measured finesse $F_{\rm cav}=180$, the free spectral 
range $\nu_{\rm FSR}=790$ MHz and the spectrum analyzer frequency 
$\Omega/2\pi=1.1$ MHz \cite{Ou}.

In the teleportation experiment of Ref.~\onlinecite{Furu}, the teleported 
states described fields at modulation frequency $\Omega/2\pi=2.9$ MHz
within a bandwidth $\pm\Delta\Omega/2\pi=30$ kHz. 
Due to technical noise at low modulation frequencies, the nonclassical
fidelity was achieved at these higher frequencies $\Omega$.
The amount of squeezing at these frequencies was about 3 dB.
The spectrum of the fidelities from Eq.~(\ref{fidspec2}) is shown
in Fig.~7 for different $\epsilon$ values.

\section{Summary and conclusions}

We have presented the broadband theory for quantum teleportation
using squeezed-state entanglement. Our scheme allows the broadband 
transmission of nonorthogonal quantum states. We have discussed
various criteria determining the boundary between classical
teleportation (i.e., measuring the state to be transmitted as
well as quantum theory permits and classically conveying the 
results) and quantum teleportation (i.e., using entanglement for the
state transfer). Depending on the set of input states, different 
criteria can be applied that are best met with the optimum gain used
by Bob for the phase-space displacements of his EPR beam.
Given an alphabet of arbitrary Gaussian states with unknown coherent
amplitudes, on average, the optimum teleportation fidelity is attained
with unit gain at all relevant frequencies. Optimal teleportation
of an entangled state (entanglement swapping) requires a 
squeezing-dependent, and hence frequency-dependent, nonunit gain.
Effectively, also with optimum gain, the bandwidth of entanglement
becomes smaller after entanglement swapping compared to the bandwidth
of entanglement of the initial states, as the quality of the entanglement
deteriorates at each frequency for finite squeezing.
 
In the particular case of the NOPA as the entanglement source,
the best quantum teleportation occurs in the frequency regime close to the 
center frequency (half the NOPA's pump frequency).
In general, a suitable EPR source for broadband teleportation can be 
obtained by combining two independent broadband squeezed states
at a beamsplitter (actually, even one squeezed state split at a 
beamsplitter is sufficient to create entanglement for quantum 
teleportation \cite{PvL4,PvL}). Provided ideal Bell detection, unit-gain
teleportation will then in general produce an excess noise in each 
teleported quadrature of twice the squeezing spectrum of the quiet 
quadrature in the corresponding broadband squeezed state (for the NOPA, 
cavity loss appears in the squeezing spectrum). 
Thus, good broadband teleportation requires good broadband squeezing.
However, the entanglement source's squeezing 
spectrum for its quiet quadrature need not be a minimum near the center 
frequency ($\Omega=0$) as for the optical parametric oscillator. In general, 
it might have large excess noise there and be quiet at $\Omega\neq 0$ as 
for four-wave mixing in a cavity \cite{Slush}.
The spectral range to be teleported $\Delta\Omega$ always should be in 
the ``quiet region'' of the squeezing spectrum.

The scheme presented here allows very efficient teleportation of
broadband quantum states: the quantum state at the input 
(a coherent, a squeezed, an entangled or any other state), describing
the input field at modulation frequency $\Omega$ within a bandwidth
$\Delta\Omega$, is teleported on each and every trial (where the
duration of a single trial is given by the inverse-bandwidth time
$1/\Delta\Omega$). Every inverse-bandwidth time, a quantum state is 
teleported with nonclassical fidelity or previously unentangled fields 
become entangled. Also the output of entanglement swapping can therefore
be used for efficient quantum teleportation, succeeding every 
inverse-bandwidth time.

In contrast, the discrete-variable schemes involving weak down conversion
enable only relatively rare transfers of quantum states.
For the experiment of 
Ref.~\onlinecite{Bou}, a fourfold coincidence (i.e., ``successful'' 
teleportation \cite{Sam3}) at a rate of 1/40 Hz and a UV pulse rate of 80 
MHz \cite{Weinf} yield an overall efficiency of $3\times 10^{-10}$ (events 
per pulse). Note that due to filtering and collection difficulties the 
photodetectors in this experiment operated with an effective efficiency of 
10\% \cite{Weinf}.

The theory presented in this paper applies to the experiment of 
Ref.~\onlinecite{Furu} where coherent states were teleported
using the entanglement built from two squeezed fields generated
via degenerate down conversion.
The experimentally determined fidelity in this 
experiment was $F=0.58\pm 0.02$ (this fidelity was 
achieved at higher frequencies $\Omega\neq 0$ due to technical noise 
at low modulation frequencies) which proved the quantum nature of 
the teleportation process by exceeding the classical limit 
$F\leq\case{1}{2}$.    
Our analysis was also intended to provide the theoretical foundation
for the teleportation of quantum states that are more nonclassical
than coherent states, e.g., squeezed states or, in particular,
entangled states (two-mode squeezed states).
This is yet to be realized in the laboratory.
 
\acknowledgments
 
The authors would like to thank C.\ M.\ Caves for helpful suggestions.
P.v.L. thanks T.\ C.\ Ralph, H.\ Weinfurter, and A.\ Sizmann for their help.
This work was supported by EPSRC Grant No.~GR/L91344.
P.v.L. was funded in part by a ``DAAD Doktorandenstipendium im
Rahmen des gemeinsamen Hochschulsonderprogramms III von Bund und
L\"{a}ndern." H.J.K. is supported by DARPA via the QUIC Institute which
is administered by ARO, by the National Science Foundation, and by
the Office of Naval Research.

\newpage 
${~}$
\vskip 3truein

\begin{figure}[htb]
\begin{psfrags}
     \psfrag{1}{\Large ``in''}
     \psfrag{2}{\Large 1}
     \psfrag{3}{\Large 2}
     \psfrag{EPR}{\Large{\bf EPR}}
     \psfrag{tel}{\Large ``tel''}
     \psfrag{Bob}{\Large{\bf Bob}}
     \psfrag{Alice}{\Large{\bf Alice}}
     \psfrag{x}{\Large ${\rm D}_x$}
     \psfrag{p}{\Large ${\rm D}_p$}
     \psfrag{a}{\Large u}
     \psfrag{b}{\Large v} 
     \epsfxsize=6in
     \epsfbox[-80 -80 450 300]{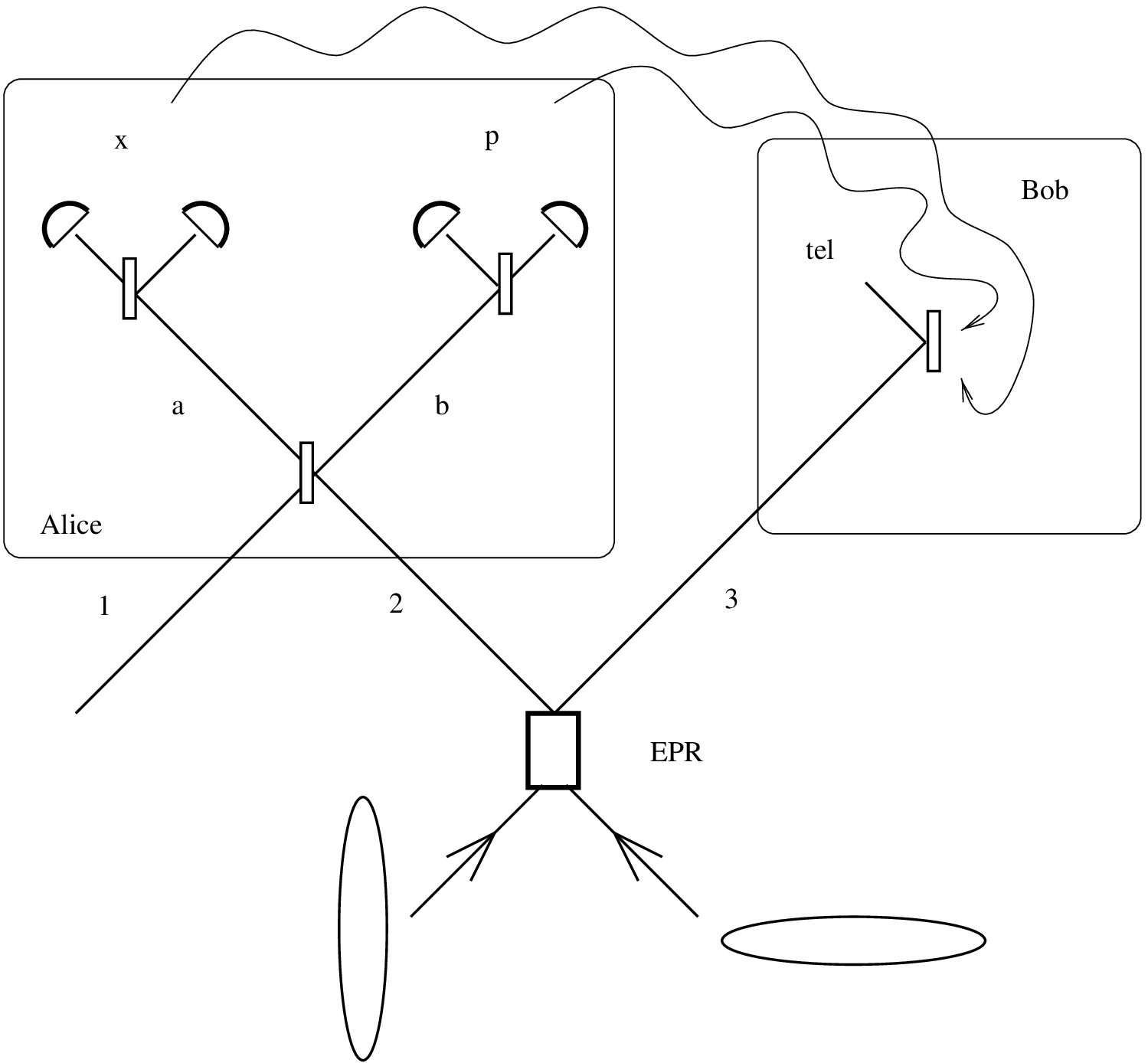}
\end{psfrags}
\caption{Teleportation of a single mode of the electromagnetic field as 
in Ref.~\protect\cite{Sam}. Alice and Bob share the entangled state of 
modes 1 and 2. Alice combines the mode ``in'' to be teleported with
her half of the EPR state at a beam splitter. The homodyne 
detectors ${\rm D}_x$ and ${\rm D}_p$ yield classical photocurrents 
for the quadratures $x_{\rm u}$ and $p_{\rm v}$, respectively. 
Bob performs phase-space displacements of his half of the EPR state
depending on Alice's classical results.}
\label{fig1}
\end{figure}

\newpage 
${~}$
\vskip 3truein

\begin{figure}[htb]
\begin{psfrags}
     \psfrag{1}{\Large ``in''}
     \psfrag{2}{\Large 1}
     \psfrag{3}{\Large 2}
     \psfrag{EPR}{\Large{\bf EPR}}
     \psfrag{tel}{\Large ``tel''}
     \psfrag{Bob}{\Large{\bf Bob}}
     \psfrag{Alice}{\Large{\bf Alice}}
     \psfrag{Victor}{\Large{\bf Victor}}
     \psfrag{x}{\Large ${\rm D}_x$}
     \psfrag{p}{\Large ${\rm D}_p$}
     \psfrag{a}{\Large u}
     \psfrag{b}{\Large v} 
     \epsfxsize=6in
     \epsfbox[-80 -80 450 300]{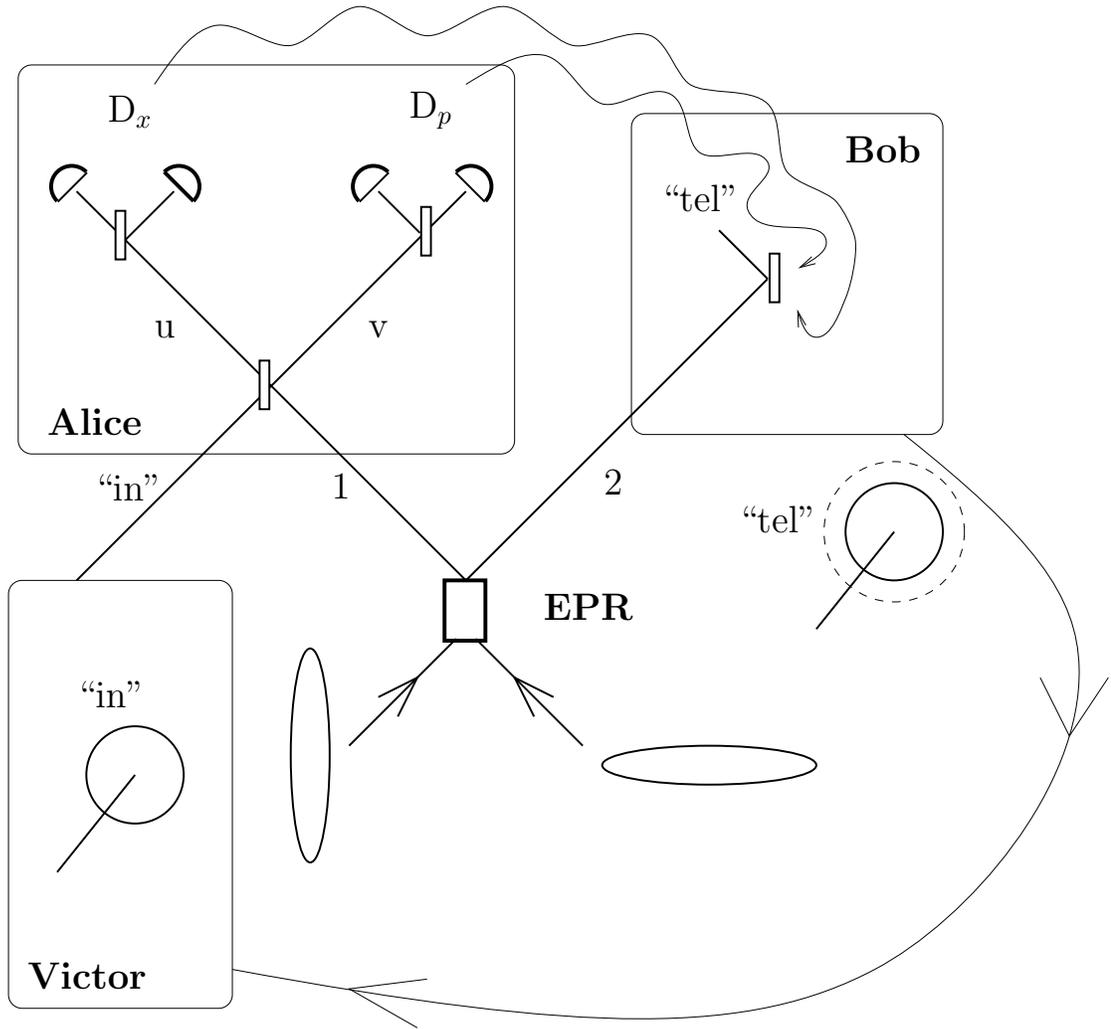}
\end{psfrags}
\caption{Verification of quantum teleportation. The verifier ``Victor''
is independent of Alice and Bob. Victor prepares the input states
which are known to him, but unknown to Alice and Bob. After a supposed
quantum teleportation from Alice to Bob, the teleported states are given
back to Victor. Due to his knowledge of the input states, 
Victor can compare the teleported states with the input states.}
\label{fig2}
\end{figure}

\newpage
${~}$
\vskip 3truein

\begin{figure}[htb]
\begin{center}
\begin{psfrags}
     \psfrag{Alice}{\Huge \bf Alice}
     \psfrag{Bob}{\Huge \bf Bob}
     \psfrag{Claire}{\Huge \bf $\;$Claire}
     \psfrag{1}{\Huge 1}
     \psfrag{2}{\Huge 2}
     \psfrag{3}{\Huge 3}
     \psfrag{4}{\Huge 4}
     \psfrag{x}{\Huge ${\rm D}_x~~~~~$}
     \psfrag{p}{\Huge ${\rm D}_p~~~~~$}
     \psfrag{u}{\Huge u}
     \psfrag{v}{\Huge v} 
     \epsfxsize=4.8in
     \epsfbox[-60 -60 400 320]{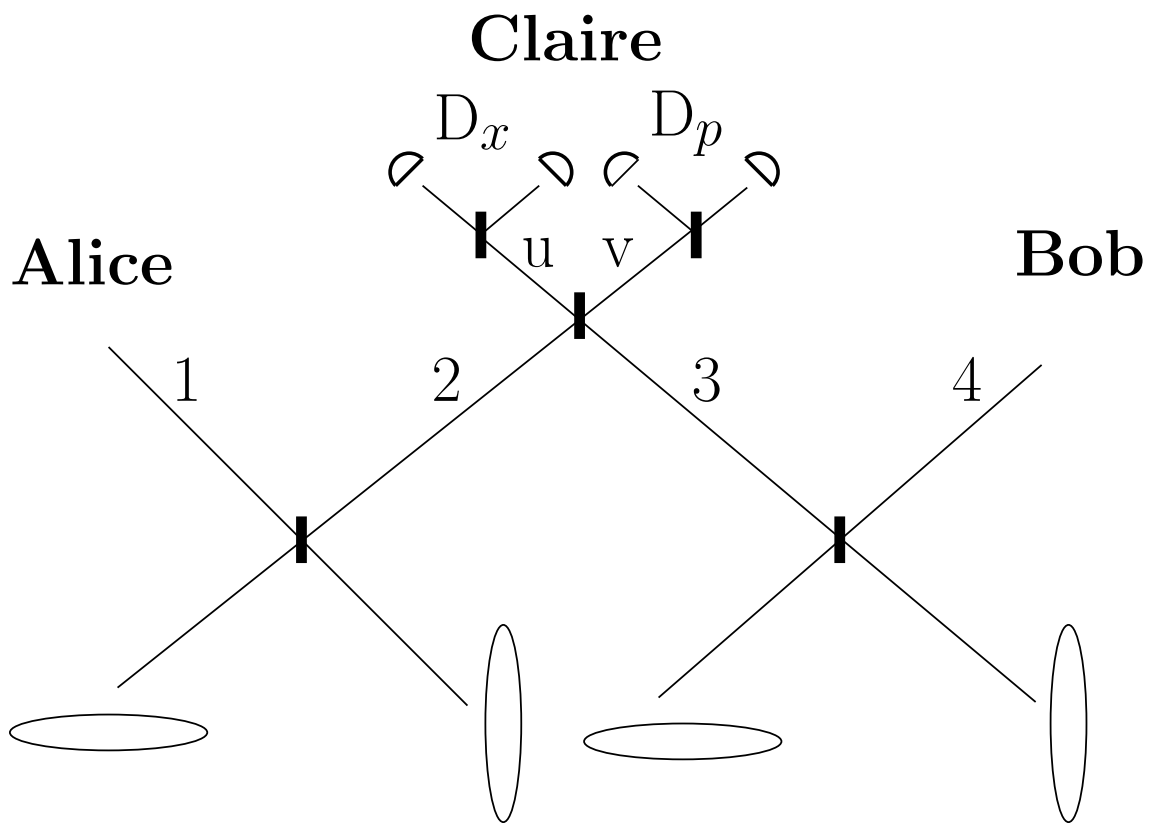}
\end{psfrags}
\end{center}
\caption{Entanglement swapping using the two entangled two-mode squeezed 
vacuum states of modes 1 and 2 (shared by Alice and Claire) and of 
modes 3 and 4 (shared by Claire and Bob) as in Ref.~\protect\cite{PvL}.} 
\label{fig3}
\end{figure}

\newpage
${~}$
\vskip 3truein

\begin{figure}[htb]
\begin{center}
\begin{psfrags}
     \psfrag{c1}{\Large $\hat{c}_1^{(0)}$}
     \psfrag{c2}{\Large $\hat{c}_2^{(0)}$}
     \psfrag{a1}{\Large $\hat{a}_1$}
     \psfrag{a2}{\Large $\hat{a}_2$}
     \psfrag{b1i}{\Large $\hat{b}_1^{(0)}$}
     \psfrag{b2i}{\Large $\hat{b}_2^{(0)}$}
     \psfrag{b1}{\Large $\hat{b}_1$}
     \psfrag{b2}{\Large $\hat{b}_2$}
     \psfrag{g}{\Large $\gamma$}
     \psfrag{r}{\Large $\rho$}
     \psfrag{chi}{\Large $\chi^{(2)}$} 
     \epsfxsize=6in
     \epsfbox[-10 -20 300 198]{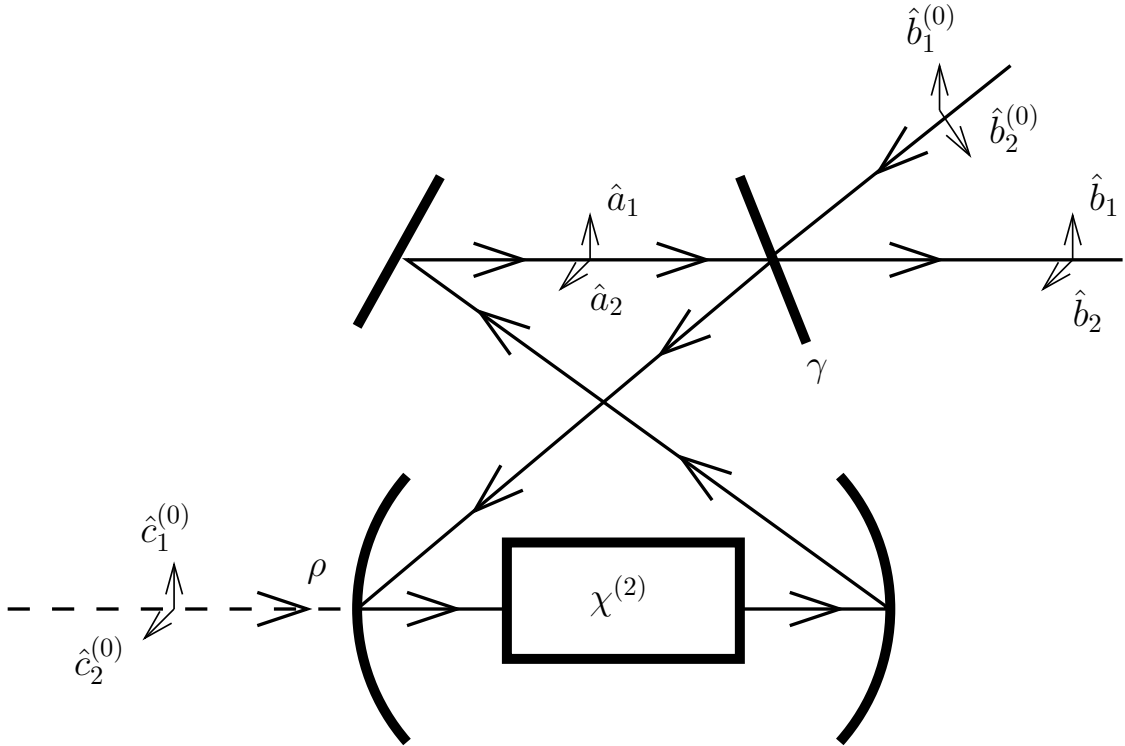}
\end{psfrags}
\end{center}
\caption{The NOPA as in Ref.~\protect\cite{Ou}. The two cavity 
modes $\hat{a}_1$ and $\hat{a}_2$ interact due to the nonlinear 
$\chi^{(2)}$ medium. The modes $\hat{b}_1^{(0)}$ and $\hat{b}_2^{(0)}$ 
are the external vacuum input modes, $\hat{b}_1$ and $\hat{b}_2$ are 
the external output modes, $\hat{c}_1^{(0)}$ and $\hat{c}_2^{(0)}$ are 
the vacuum modes due to cavity losses, $\gamma$ is a damping rate and 
$\rho$ is a loss parameter of the cavity.} 
\label{fig4}
\end{figure}

\newpage
${~}$
\vskip 3truein

\begin{figure}[htb]
\begin{center}
\begin{psfrags}
     \psfrag{F}[cb]{\Huge $F~~~$}
     \psfrag{frequency}[c]{{\Huge ~~~~~~~~~~~frequency} {\Large $\pm$}
{\Huge $\omega$}}
     \psfrag{e=0.1}[c]{\Large $~~~~~~~~~~0.1$}
     \psfrag{e=0.2}[c]{\Large $~~~~~~~~~~0.2$}
     \psfrag{e=0.4}[c]{\Large $~~~~~~~~~~0.4$}
     \psfrag{e=0.6}[c]{\Large $~~~~~~~~~~0.6$}
     \psfrag{e=1}[c]{\Large $~~~~~1$}
     \psfrag{classical}{\Huge \bf ~classical}
     \psfrag{quantum}{\Huge \bf quantum}
     \epsfxsize=5.8in
     \epsfbox[-60 0 400 320]{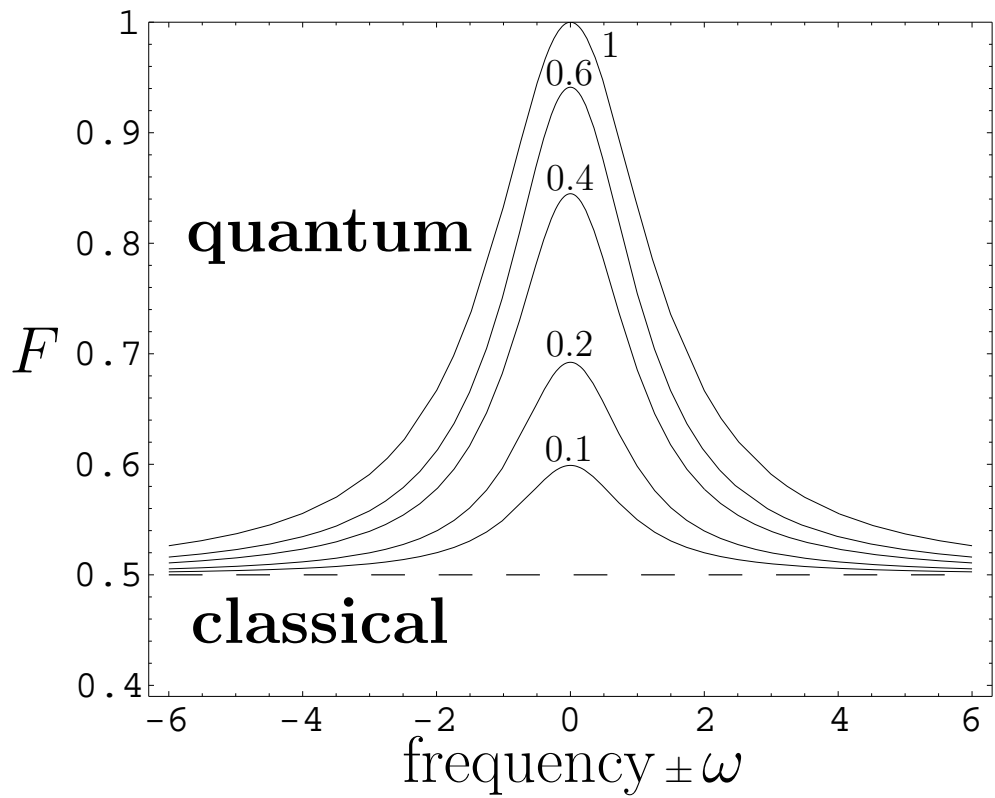}
\end{psfrags}
\end{center}
\caption{Fidelity spectrum of coherent-state teleportation using 
entanglement from the NOPA. The fidelities here are 
functions of the normalized modulation frequency $\pm\omega$ for 
different parameter $\epsilon$ ($=0.1$, $0.2$, $0.4$, $0.6$, and $1$).}
\label{fig5}
\end{figure}

\newpage
${~}$
\vskip 3truein

\begin{figure}[htb]
\begin{center}
\begin{psfrags}
     \psfrag{F}[cb]{\Huge $F~~~$}
     \psfrag{frequency}[c]{{\Huge ~~~~~~~~~~~frequency} {\Large $\pm$}
{\Huge $\omega$}}
     \psfrag{e=0.1}[c]{\Large $~~~~~~~~~~0.1$}
     \psfrag{e=0.2}[c]{\Large $~~~~~~~~~~0.2$}
     \psfrag{e=0.4}[c]{\Large $~~~~~~~~~~0.4$}
     \psfrag{e=0.6}[c]{\Large $~~~~~~~~~~0.6$}
     \psfrag{e=1}[c]{\Large $~~~~~1$}
     \psfrag{classical}{\Huge \bf ~classical}
     \psfrag{quantum}{\Huge \bf quantum}
     \epsfxsize=5.8in
     \epsfbox[-60 0 400 320]{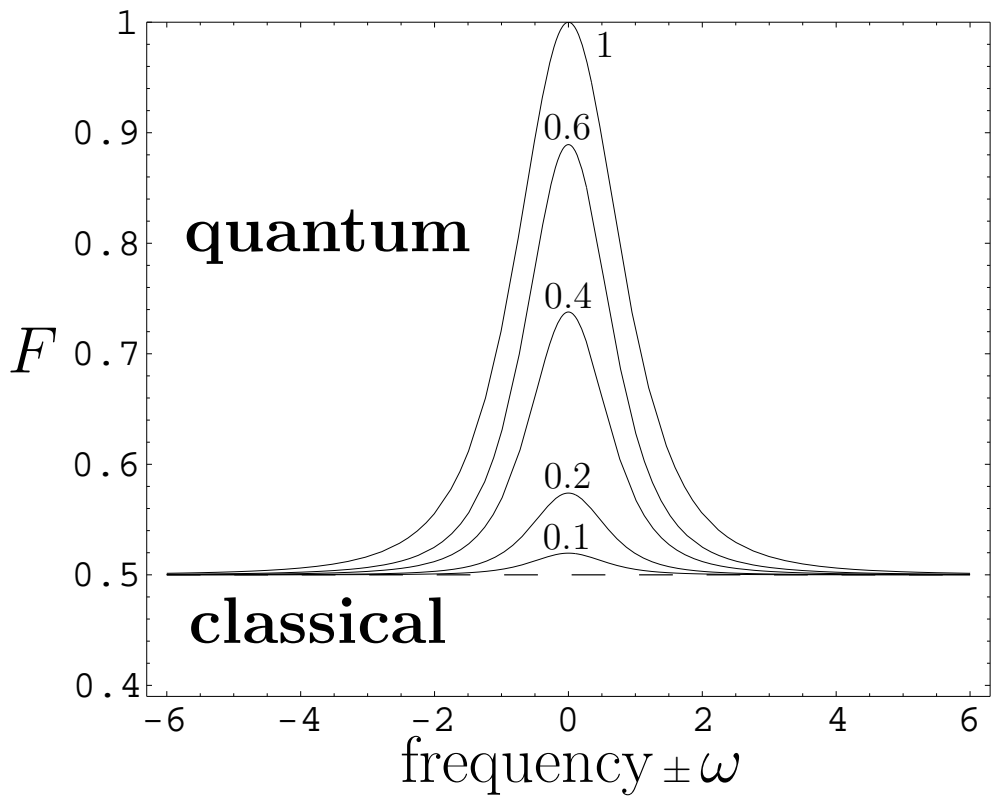}
\end{psfrags}
\end{center}
\caption{Fidelity spectrum of coherent-state teleportation using the output 
of entanglement swapping with two equally squeezed (entangled) NOPA's.
The fidelities here are functions of the normalized modulation 
frequency $\pm\omega$ for different parameter $\epsilon$ 
($=0.1$, $0.2$, $0.4$, $0.6$, and $1$).}
\label{fig6}
\end{figure}

\newpage
${~}$
\vskip 3truein

\begin{figure}[htb]
\begin{center}
\begin{psfrags}
     \psfrag{F}[cb]{\Huge $F~~~$}
     \psfrag{frequency}[c]{{\Huge ~~~~~~~~~~~frequency} {\Large $\pm$}
{\Huge $\omega$}}
     \psfrag{e=0.1}[c]{\Large $~~~~~~~~~~0.1$}
     \psfrag{e=0.2}[c]{\Large $~~~~~~~~~~0.2$}
     \psfrag{e=0.4}[c]{\Large $~~~~~~~~~~0.4$}
     \psfrag{e=0.6}[c]{\Large $~~~~~~~~~~0.6$}
     \psfrag{e=1}[c]{\Large $~~~~~1$}
     \psfrag{classical}{\Huge \bf ~classical}
     \psfrag{quantum}{\Huge \bf quantum}
     \epsfxsize=5.8in
     \epsfbox[-60 0 400 320]{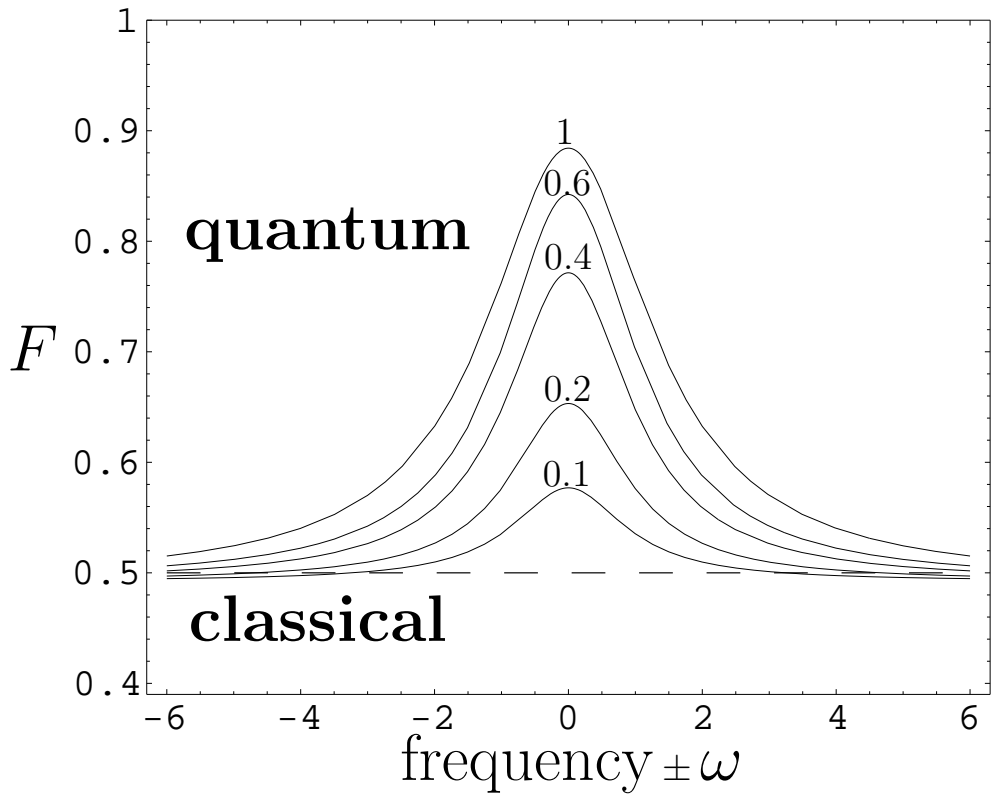}
\end{psfrags}
\end{center}
\caption{Fidelity spectrum of coherent-state teleportation using 
entanglement from the NOPA. The fidelities here are 
functions of the normalized modulation frequency $\pm\omega$ for 
different parameter $\epsilon$ ($=0.1$, $0.2$, $0.4$, $0.6$, and $1$).
Bell detector efficiencies $\eta^2=0.97$ and cavity losses with 
$\beta=0.9$ have been included here.}
\label{fig7}
\end{figure}

\end{document}